\documentclass[lettersize,journal]{IEEEtran}
\usepackage{amsmath,amsfonts}
\usepackage[algo2e,ruled]{algorithm2e}
\usepackage{array}
\usepackage{textcomp}
\usepackage{stfloats}
\usepackage{url}
\usepackage{verbatim}
\usepackage[inline]{enumitem}
\usepackage{graphicx}
\usepackage{colortbl}
\usepackage{nicefrac}
\usepackage{float}
\usepackage{subcaption}
\usepackage{pifont}
\usepackage{dsfont}
\usepackage[numbers,sort]{natbib}
\usepackage{makecell}

\usepackage{tikz}
\usepackage{pgf}
\usepackage{pgfplots}
\usepackage{pgfplotstable}
\usepackage{arydshln}

\pgfplotsset{compat=1.18}
\usepgfplotslibrary{external} 
\usepgfplotslibrary{fillbetween} 
\usepgfplotslibrary{groupplots} 
\usetikzlibrary{arrows.meta}
\usetikzlibrary{decorations.text}    
\usetikzlibrary{math}

\makeatletter
\newcommand{\removelatexerror}{\let\@latex@error\@gobble}
\makeatother

\newcommand{\algorithmstyle}[1]{\renewcommand{\algocf@style}{#1}}

\SetKwComment{Comment}{$\triangleright$\ }{}

\makeatletter
\newcommand{\algorithmfootnote}[2][\footnotesize]{%
  \let\old@algocf@finish\@algocf@finish
  \def\@algocf@finish{\old@algocf@finish
    \leavevmode\rlap{\begin{minipage}{\linewidth}
    #1#2
    \end{minipage}}%
  }%
}
\makeatother

\newcommand{\nocontentsline}[3]{}
\newcommand{\tocless}[2]{\bgroup\let\addcontentsline=\nocontentsline#1{#2}\egroup}

\usepackage[backgroundcolor=blue!20!white,bordercolor=white]{todonotes}

\definecolor{cherry1}{rgb}{0.215686, 0.215686, 0.215686}
\definecolor{cherry2}{rgb}{0.563899, 0.155919, 0.156577}
\definecolor{cherry3}{rgb}{0.747389, 0.178584, 0.180272}
\definecolor{cherry4}{rgb}{0.836168, 0.264453, 0.26819}
\definecolor{cherry5}{rgb}{0.880144, 0.397868, 0.404399}
\definecolor{cherry6}{rgb}{0.911942, 0.567676, 0.576412}

\definecolor{ibmred1}{rgb}{0.643137,0.086274,0.30588}
\definecolor{ibmred2}{rgb}{0.925490,0.325490,0.545098}
\definecolor{ibmred3}{rgb}{0.996078,0.623529,0.764705}
\definecolor{ibmpurple1}{rgb}{0.427451,0.1921569,0.7843137}
\definecolor{ibmpurple2}{rgb}{0.6588235,0.4352941,0.9960784}
\definecolor{ibmpurple3}{rgb}{0.8156863,0.6901961,1.0}
\definecolor{ibmblue1}{rgb}{0.0,0.345098,0.627451}
\definecolor{ibmblue2}{rgb}{0.06666667,0.5764706,0.9058824}
\definecolor{ibmblue3}{rgb}{0.4235294,0.7921569,0.9960784}

\usepackage{letltxmacro}
\usepackage{xparse}
\makeatletter
\LetLtxMacro\ieeetran@appendix\appendix
\AtBeginDocument{%
\RenewDocumentCommand{\appendix}{o}{%
  \IfValueTF{#1}{%
    \ieeetran@appendix[#1]%
  }{%
    \ieeetran@appendix%
  }%
}
}
\makeatother

\usepackage{amssymb}

\usepackage{amsmath,amsthm}
\usepackage{mathtools}
\usepackage{iftex}
\usepackage{etoolbox}

\ifxetex
  \usepackage{microtype}

  \newcommand{\mathmat}[1]{\mathbfit{#1}}
  \newcommand{\mathvec}[1]{\mathbfit{#1}}
  \newcommand{\mathvecgreek}[1]{\mathbfit{#1}}
  \newcommand{\mathmatgreek}[1]{\mathbfit{#1}}
  
  \newcommand{\boldupright}[1]{\symbfup{#1}}

  \newcommand{\mathrv}[1]{\mathsfit{#1}}
  \newcommand{\mathrvvec}[1]{\mathbfsfit{#1}}
  \newcommand{\mathrvgreek}[1]{\mathsfit{#1}}
  \newcommand{\mathrvvecgreek}[1]{\mathbfsfit{#1}}

  \usepackage{fontspec}
\else
  \usepackage{textcomp}

  \newcommand{\mathmat}[1]{\boldsymbol{#1}}
  \newcommand{\mathvec}[1]{\boldsymbol{#1}}
  \newcommand{\mathvecgreek}[1]{\mathit{\boldsymbol{#1}}}
  \newcommand{\mathmatgreek}[1]{\mathit{\boldsymbol{#1}}}
  
  \newcommand{\boldupright}[1]{\mathbf{#1}}

  \newcommand{\mathrv}[1]{\mathit{#1}}
  \newcommand{\mathrvvec}[1]{\boldsymbol{#1}}
  \newcommand{\mathrvgreek}[1]{\mathit{\boldsymbol{#1}}}
  \newcommand{\mathrvvecgreek}[1]{\mathit{\boldsymbol{#1}}}
\fi

\usepackage{booktabs,threeparttable,arydshln}
\usepackage{multirow}
\usepackage{nicematrix}

\usepackage[algo2e, ruled]{algorithm2e}

\usepackage{pgfplots}
\pgfkeys{/pgfplots/tuftelike/.style={
  semithick,
  tick style={major tick length=4pt,semithick,black},
  separate axis lines,
  axis x line*=bottom,
  axis x line shift=2pt,
  xlabel shift=-2pt,
  axis y line*=left,
  axis y line shift=2pt,
  ylabel shift=-2pt}}
  
\usepackage{proof-at-the-end}

\usepackage{mdframed}
\mdfdefinestyle {mdleftbarcoloredstyle} {%
    leftline          = true,%
    rightline         = false,%
    topline           = false,%
    bottomline        = false,%
    linewidth         = 2.5pt,%
    backgroundcolor   = gray!10,%
    linecolor         = gray,%
    innertopmargin    = 0.0ex,%
    innerbottommargin = 0.7ex,%
    innerleftmargin   = 1.ex,%
    ntheorem          = false,%
}

\mdfdefinestyle {coloredstyle} {%
    leftline          = false,%
    rightline         = false,%
    topline           = false,%
    bottomline        = false,%
    backgroundcolor   = gray!10,%
    innertopmargin    = 0.0ex,%
    innerbottommargin = 0.7ex,%
    innerleftmargin   = 1.ex,%
}

\declaretheoremstyle[
    spaceabove    = \parsep,
    spacebelow    = \parsep,
    bodyfont      = \normalfont\itshape,
]{theoremsty}

\declaretheorem[name=Proposition,style=theoremsty]{proposition}

\declaretheorem[name=Theorem,    style=theoremsty, mdframed={style = coloredstyle}, numbered=no]{theorem*}
\declaretheorem[name=Proposition,style=theoremsty, mdframed={style = coloredstyle}, numbered=no]{proposition*}
\declaretheorem[name=Corollary,  style=theoremsty, mdframed={style = coloredstyle}, numbered=no]{corollary*}
\declaretheorem[name=Lemma,      style=theoremsty, mdframed={style = coloredstyle}, numbered=no]{lemma*}

\declaretheorem[name=Open Problem,style=theoremsty, mdframed={style = coloredstyle}, numbered=no]{openproblem*}
\declaretheorem[name=Conjecture,  style=theoremsty, mdframed={style = coloredstyle}, numbered=no]{conjecture*}

\declaretheoremstyle[
    spaceabove=\parsep,
    spacebelow=\parsep,
    bodyfont=\normalfont,
]{normalsty}
\declaretheorem[name=Remark,      style=normalsty]{remark}
\declaretheorem[name=Definition,  style=normalsty]{definition}

\declaretheorem[name=Remark,      style=normalsty, numbered=no]{remark*}
\declaretheorem[name=Definition,  style=normalsty, mdframed={style = coloredstyle}, numbered=no]{definition*}
\declaretheorem[name=Assumption,  style=normalsty, mdframed={style = coloredstyle}, numbered=no]{assumption*}

\makeatletter
\providecommand\IfFormatAtLeastTF{\@ifl@t@r\fmtversion}
\IfFormatAtLeastTF{2020/10/02}{%
  \usepackage[bookmarksopen=true,bookmarks=true,colorlinks=true,breaklinks]{hyperref}

  \usepackage[capitalise, nameinlink]{cleveref}
}{%
  \usepackage{hyperref}
  \usepackage[capitalise, nameinlink]{cleveref}%
}
\makeatother

\definecolor{linkcolor}{HTML}{6929C4}
\definecolor{citecolor}{HTML}{0043CE}
\hypersetup{colorlinks={true},linkcolor=linkcolor,citecolor=citecolor}

\crefname{assumption}{Assumption}{Assumption}
\crefname{condition}{Condition}{Condition}

\crefname{framedtheorem}{Thm.}{Thms.}
\crefname{framedproposition}{Prop.}{Props.}
\crefname{framedlemma}{Lem.}{Lems.}
\crefname{theorem}{Thm.}{Thms.}
\crefname{proposition}{Prop.}{Props.}
\crefname{lemma}{Lem.}{Lems.}

\makeatletter
\def\adl@drawiv#1#2#3{%
        \hskip.5\tabcolsep
        \xleaders#3{#2.5\@tempdimb #1{1}#2.5\@tempdimb}%
                #2\z@ plus1fil minus1fil\relax
        \hskip.5\tabcolsep}
\newcommand{\cdashlinelr}[1]{%
  \noalign{\vskip\aboverulesep
           \global\let\@dashdrawstore\adl@draw
           \global\let\adl@draw\adl@drawiv}
  \cdashline{#1}
  \noalign{\global\let\adl@draw\@dashdrawstore
           \vskip\belowrulesep}}
\makeatother

\pgfkeys{/prAtEnd/global custom defaults/.style={
    end,
  }
}

\DeclareMathOperator*{\argmin}{arg\,min} 

\newcommand*\xbar[1]{%
  \hbox{%
    \vbox{%
      \hrule height 0.6pt 
      \kern0.33ex
      \hbox{%
        \kern-0.1em
        \ensuremath{#1}%
        \kern-0.1em
      }%
    }%
  }%
}

\newcommand{\norm}[1]{{\left\lVert #1 \right\rVert}}
\newcommand{\abs}[1]{{\left| #1 \right|}}

\def\do#1{\csdef{v#1}{\mathvec{#1}}}
\docsvlist{a,b,c,d,e,f,g,h,i,j,k,l,m,n,o,p,q,r,s,t,u,v,w,x,y,z}
\docsvlist{A,B,C,D,E,F,G,H,I,J,K,L,M,N,O,P,Q,R,S,T,U,V,W,X,Y,Z}

\def\do#1{\csdef{v#1}{\mathvecgreek{\csname#1\endcsname}}}
\docsvlist{alpha,beta,gamma,delta,epsilon,zeta,eta,theta,iota,kappa,lambda,mu,nu,xi,omikron,pi,rho,sigma,tau,upsilon,phi,chi,psi,omega}

\def\do#1{\csdef{rv#1}{\mathrv{#1}}}
\docsvlist{a,b,c,d,e,f,g,h,i,j,k,l,m,n,o,p,q,r,s,t,u,v,w,x,y,z}
\docsvlist{A,B,C,D,E,F,G,H,I,J,K,L,M,N,O,P,Q,R,S,T,U,V,W,X,Y,Z}

\def\do#1{\csdef{rv#1}{\mathrvgreek{\csname#1\endcsname}}}
\docsvlist{alpha,beta,gamma,delta,epsilon,zeta,eta,theta,iota,kappa,lambda,mu,nu,xi,omikron,pi,rho,sigma,tau,upsilon,phi,chi,psi,omega}

\def\do#1{\csdef{rvv#1}{\mathrvvec{#1}}}
\docsvlist{a,b,c,d,e,f,g,h,i,j,k,l,m,n,o,p,q,r,s,t,u,v,w,x,y,z}

\def\do#1{\csdef{rvv#1}{\mathrvvecgreek{\csname#1\endcsname}}}
\docsvlist{alpha,beta,gamma,delta,epsilon,zeta,eta,theta,iota,kappa,lambda,mu,nu,xi,omikron,pi,rho,sigma,tau,upsilon,phi,chi,psi,omega}

\def\do#1{\csdef{rvm#1}{\mathrvvecgreek{#1}}}
\docsvlist{A,B,C,D,E,F,G,H,I,J,K,L,M,N,O,P,Q,R,S,T,U,V,W,X,Y,Z}

\def\do#1{\csdef{m#1}{\mathmat{#1}}}
\docsvlist{A,B,C,D,E,F,G,H,I,J,K,L,M,N,O,P,Q,R,S,T,U,V,W,X,Y,Z}

\def\do#1{\csdef{m#1}{\mathmatgreek{\csname#1\endcsname}}}
\docsvlist{Sigma,Lambda,Phi,Gamma,Theta,Delta}

\newcommand{\herm}{\dagger}

\begin{document}
\bstctlcite{IEEEexample:BSTcontrol}

\title{Bayesian Wideband Signal Detection via\\Source Signal Marginalization and RJMCMC}

\author{Kyurae Kim,~\IEEEmembership{Member,~IEEE,}
Philip T. Clemson, \\
James P. Reilly,~\IEEEmembership{Life Member,~IEEE,}
Jason F. Ralph,
Simon Maskell,~\IEEEmembership{Member,~IEEE}
\thanks{K. Kim was with the University of Liverpool, Liverpool, United Kingdom, but is now with the University of Pennsylvania, Philadelphia, Pennsylvania, United States (e-mail: kyrkim@seas.upenn.edu).}
\thanks{P. T. Clemson was with the University of Liverpool, Liverpool, United Kingdom, but is now with Lancaster University, Lancaster, United Kingdom. (e-mail: p.t.clemson1@lancaster.ac.uk)}
 \thanks{J. Reilly is with McMaster University, Hamilton, Ontario, Canada (e-mail: reillyj@mcmaster.ca).}
\thanks{J. F. Ralph and S. Maskell are with the University of Liverpool, Liverpool, United Kingdom (e-mail: \{jfralph, S.Maskell\}@liverpool.ac.uk).}
}



\makeatletter
\def\ps@IEEEtitlepagestyle{
  \def\@oddfoot{\mycopyrightnotice}
  \def\@evenfoot{}
}
\def\mycopyrightnotice{
  \begin{minipage}{\textwidth}
    \centering\scriptsize
    \copyright 2026 IEEE. Personal use of this material is permitted.  Permission from IEEE must be obtained for all other uses, in any current or future media, including reprinting/republishing this material for advertising or promotional purposes, creating new collective works, for resale or redistribution to servers or lists, or reuse of any copyrighted component of this work in other works.
  \end{minipage}
}
\makeatother

\maketitle

\begin{abstract}
Consider an array receiving unknown wideband signals from an unknown number of sources $k$.
Wideband signals can occupy arbitrarily wide bandwidths, rendering demodulation-based approaches inapplicable, a common situation in settings involving acoustic signals (\textit{e.g.}, passive sonar, seismology).
Here, we aim to determine $k$ given $N$ noisy array-valued measurements, a task known as the ``detection problem,'' for which Bayesian model comparison is a common approach.
To render Bayesian inference tractable, it is typically necessary to marginalize the source signals.
Unfortunately, for wideband signals, naive marginalization has an unaffordable time complexity of \(\mathcal{O}(N^3 k^3)\).
As a result, fully Bayesian signal detection has yet to be demonstrated in wideband settings.
In this work, we propose a wideband signal model that allows for computationally tractable marginalization of the source signals.
We begin from the canonical model of linear time-invariant (LTI) signal propagation, which is then augmented into a circular convolution, all without loss of generality.
This allows for efficient computation in the frequency domain, where the resulting linear system admits a decomposition into a sparse matrix we refer to as a \textit{stripe matrix decomposition}.
Exploiting this sparsity pattern reduces the time complexity of computing the marginal likelihood to \(\mathcal{O}(N k^3)\).
These computational improvements enable efficient posterior inference via reversible-jump Markov chain Monte Carlo (RJMCMC).
In this work, we use the non-reversible extension of RJMCMC (NRJMCMC), which often achieves lower autocorrelation and faster convergence than RJMCMC.
Detection of the latent source signals can then be performed in a fully Bayesian manner using samples drawn by NRJMCMC.
We evaluate our procedure by comparing it against generalized likelihood ratio testing (GLRT) and information criteria.
\end{abstract}


\section{Introduction}
\IEEEPARstart{C}{consider} an array of sensors facing an unknown number of sources, each emitting a wideband signal that we have no knowledge of.
Here, wideband implies that the signals may occupy arbitrarily wide bandwidths or any unknown set of frequency bands, making demodulation to a narrower band inapplicable.
This setting, known as the \textit{wideband setting}, commonly arises in applications involving acoustic signals such as passive sonar~\citep{zijiantang_aliasingfree_2011,baggeroer_passive_1999}.
Here, the task of determining the number of sources from noisy measurements obtained from the array is known as the signal detection problem~\citep[\S 3.14.7]{chung_doa_2014}.
This differs from the classic direction-of-arrival (DoA) estimation problem~\citep{chung_doa_2014,vantrees_optimum_2002,krim_two_1996} in the sense that the detection problem asks for the more fundamental question of \textit{what constitutes a signal}.
In statistical terms, this corresponds to a (nested) model selection or model order determination~\citep{koivunen_model_2014,stoica_modelorder_2004,ding_model_2018}.


Under the \textit{narrowband signal model}~\citep{vantrees_optimum_2002}, the detection problem has been tackled through various approaches from information-theoretic methods~\citep{wax_spatiotemporal_1984, wax_detection_1985}, hypothesis testing~\citep{kritchman_nonparametric_2009,brcich_detection_2002,viberg_detection_1991,ottersten_exact_1993,chen_detection_1991}, and Bayesian model comparison~\citep{larocque_reversible_2002,nielsen_bayesian_2014}.
For some applications, however, the signals of interest may lie in different bands or span multiple frequency bands, meaning that narrowband methods are no longer applicable.
For these applications, wideband methods have to be applied.

Wideband signal detection methods~\citep[Ch. 14.6]{chung_doa_2014} avoid relying on the narrowband assumption by directly dealing with the wideband nature of wideband signals.
Earlier attempts tried to focus the multiple frequency bands into a single reference narrowband~\citep{wax_spatiotemporal_1984,wang_coherent_1985,diclaudio_waves_2001,valaee_wideband_1995}.
These so-called focused subspace methods were popular as they allow for the use of off-the-shelf narrowband detection methods, typically information-theoretic approaches~\citep{wax_detection_1985,stoica_modelorder_2004}, on the reference band~\citep{wang_coherent_1985}.
Unfortunately, focused subspace methods require an accurate initial guess of the number of sources and their DoAs~\citep{amirsoleimani_wideband_2020,yoon_tops_2006}, resulting in a chicken-and-egg problem that complicates their use.
Furthermore, information-theoretic methods, or ``information criteria,'' are known to achieve limited performance in the low signal-to-noise ratio (SNR), low snapshot regime~\citep{nielsen_bayesian_2014,chung_detection_2007,chen_detection_1991}.

Most of the issues with focused subspace approaches stem from the fundamental concept of collapsing multiple frequency bins into one.
In contrast, maximum likelihood (ML;~\citep[\S 3.14.6.1]{chung_doa_2014}, \citep{schweppe_sensorarray_1968}) approaches rigorously model wideband signals and are known to outperform subspace methods ``when the signal-to-noise ratio is small, or alternatively, when the number of samples (``snapshots'') is small''~\citep{ziskind_maximum_1988}.
Moreover, ML methods enable detection through generalized likelihood ratio testing (GLRT; \citep[\S 3.02.3.2]{koivunen_model_2014}).
That is, detection is now viewed as a multiple hypothesis testing problem where, given the null hypothesis \(H_k\) that \(k\) targets are present, the alternative composite hypothesis \(H_{k+1}\) that at least \(k + 1\) targets are present is tested for each \(k \in \{0, \ldots, k_{\mathrm{max}} \}\) (\(k_{\mathrm{max}}\) is the assumed maximum number of targets).
For wideband signals with additive white Gaussian noise, test statistics have been derived in~\citep{boehme_statistical_1995}, 
where bootstrapping~\citep{zoubir_bootstrap_2001} has been used to compute the \(p\)-values needed for the tests~\citep{maiwald_multiple_1994,chung_detection_2007}.

While GLRT resolves the limitations of focused subspace approaches, it suffers from the usual limitations of hypothesis testing: They are designed to control the long-term false positive rate (Type-I error rate) at the cost of a higher false negative rate (Type-II error rate)~\citep[\S 8.3]{casella_statistical_2001}.
Therefore, when \textit{accuracy} or having a low false negative rate (``power'') is also of interest, the design principles of hypothesis testing become misaligned.
This problem is further exacerbated in signal detection, as it is typically approached as a multiple-testing problem~\citep{koivunen_model_2014}.
When controlling the Type-I error, the power of a multiple hypothesis test decreases as the number of considered hypotheses (\(k_{\mathrm{max}}\)) increases.

On the other hand, the Bayesian approach to model comparison~\citep{jeffreys_theory_1998,kass_bayes_1995} is immune to these issues~\citep{robert_bayesian_2001}.
In fact, the Bayesian approach can select the model that maximizes model selection \textit{accuracy}.
More precisely, assuming the model is well specified, Bayesian point estimators achieve the optimal average performance (Bayes risk) for any given target performance metric~\citep[Thm. 2.3.2]{robert_bayesian_2001}.
For signal processing problems, \textit{reversible jump} Markov chain Monte Carlo (RJMCMC; \citep{green_reversible_1995,hastie_model_2012}), also known as transdimensional MCMC~\citep{green_transdimensional_2003,sisson_transdimensional_2005}, often enables inference of the model posterior and their corresponding parameter posteriors all at once, and have been successfully applied to various signal processing problems ~\citep{andrieu_joint_1999,copsey_bayesian_2002,davy_bayesian_2006,eches_estimating_2010,davy_classification_2002,liu_bayesian_2022,amrouche_efficient_2022}, including narrowband~\citep{larocque_reversible_2002} and wideband~\citep{ng_wideband_2005} DoA estimation and detection.

Employing Bayesian model comparison in wideband signal detection, however, is not trivial.
In a wideband setting, it is natural to model the signal propagation as a linear time-invariant (LTI) system.
Consider a setup where an $M$-sensor array is receiving \(N\) measurements from \(k\) sources and the signal is propagated through an \(L\)-tap finite impulse response (FIR) filter.
Faithfully modeling this setup results in a $N M \times N k$ linear system, where there are $k$ unobserved source signals, each at least $N$-samples long.
This means $\Omega(N k)$ parameters corresponding to the source signals have to be inferred, which is expensive in a fully Bayesian setup, especially with RJMCMC.
It is thus typical to \textit{marginalize}~\citep[\S 9.3]{robert_monte_2004} the source signals~\citep{andrieu_joint_1999,larocque_reversible_2002,copsey_bayesian_2002,davy_bayesian_2006,liu_bayesian_2022,amrouche_efficient_2022}.
Unfortunately, naively doing this in a wideband setup imposes a time complexity of $\mathcal{O}(N^3 k^3)$.
As such, Ng \textit{et al.} \citep{ng_wideband_2005} proposed to substitute the latent signal posterior with plug-in estimates obtained through a recursive maximum \textit{a posteriori} (MAP) procedure.
Unfortunately, this scheme no longer fits into the Bayesian framework and ignores uncertainty on the latent source signals.
Also, we observed that their MAP procedure is unstable
when the inter-sensor delay is large
\footnote{
The MAP procedure by \citep{ng_wideband_2005} inverts sinc-based FIR delay filters. For finite-tap realization, the tails of the filters have to be truncated, introducing truncation error. 
When the inter-sensor delay is large, the main lobe of the sinc shifts closer to the truncated region, meaning that the magnitude of the truncated side lobes increase along the truncation error.
For values of $c$ corresponding to acoustic signals, we were unsuccessful in obtaining an implementation that did not diverge.
}
, making the scheme inapplicable to applications with a low propagation speed $c$ (\textit{e.g.}, sonar, seismology).
Therefore, there is room for a generally applicable and fully Bayesian approach.

In this work, we provide an alternative probabilistic model of wideband signals, which is the first to allow tractable closed-form marginalization of the latent signals.
Specifically, we model the received signal as a sum of $k$ unobserved source signals, each $N + L - 1$-sample long.
By augmenting the original FIR filter, we then show that the signal propagation can be represented as a circular convolution (\cref{section:model}).
As a result, the system matrix admits a matrix decomposition, which we refer to as a \textit{stripe decomposition}, into a sparse matrix.
This ``striped'' sparsity pattern~\citep{smolarski_diagonallystriped_2006} makes the system easy to solve via the block-LDL$^{\top}$ decomposition~\citep{fang_stability_2011}, which reduces the time complexity of computing the marginal/collapsed likelihood from \(\mathcal{O}(N^3 k^3)\) to \(\mathcal{O}(N k^3)\) (\cref{section:likelihood_computation}).
Marginalization reduces the number of parameters that need to be inferred from \(k (N + L - 1) + 2 k + 1\) to \(2 k\).
As a result, for the first time, we are able to demonstrate the fully Bayesian signal detection in a wideband setting.
Furthermore, since we directly model the propagation with an LTI system, we can directly reconstruct real-valued latent source signals, in contrast to the time--frequency spectrum-based models of previous works~\citep[\S 14.6]{chung_doa_2014}.

The overall procedure is described in \cref{section:detection_estimation_procedure}.
For inference, we leverage the non-reversible variant of RJMCMC proposed by \citet{gagnon_nonreversible_2021}, which achieves lower autocorrelation compared to conventional reversible implementations (\cref{section:rjmcmc}).
We evaluate the detection performance of our procedure in \cref{section:evaluation}.

\section{Background}
\paragraph*{Notation}
We denote vectors in bold lower case letters (\textit{e.g.}, \(\vx, \vy\)); matrices in bold capitals (\textit{e.g.}, \(\mA, \mB\)); \(\boldupright{I}_N, \boldupright{O}_N \in \mathbb{R}^{N \times N}\) are the identity and zero matrices, respectively; \(\boldupright{0}_N \in \mathbb{R}^N\) is a vector of \(N\) zeros.
\(\mA^{\herm}\) and $\mA^{*}$ are the Hermitian and complex conjugate of \(\mA\), respectively; \(\circledast\) is the discrete circular convolution.
For a discrete-time signal \(x_i[0], \ldots, x_i[N\text{-}1]\) for some \(N \in \mathbb{N}_{>0}\), omitting the index implies concatenation over the sample index such that \(\vx_i = \left(x_i[0], \ldots, x_i[N\text{-}1]\right)\).
For a collection of signals \(\vx_1, \ldots, \vx_k\), omitting the subscript or using a ``range'' as \(1\text{:}k\) implies concatenation over the collection such as \(\vx[n] = \vx_{1\text{:}k}[n] = 
\left(x_{1}[n],\, \ldots,\, x_k[n] \right)\) and \(\vx = \vx_{1\text{:}k} = \left(\vx_1, \ldots, \vx_k\right)\).
Lastly, a range of indices from 1 to \(N\) is denoted as \([N] \triangleq \{1, \ldots, N\}\).
 
\subsection{Wideband Signal Models}\label{section:doa}
We will first discuss the signal models typically used for modeling wideband signals~\citep{chung_doa_2014}. 

\vspace{0.8ex}
\paragraph{Continuous-Time Model}
Consider an \(M\) sensor array with continuous-time measurements.
The measurement, or sample, \(y_i\left(t\right)\) received by the \(i\)th sensor at time \(t \geq 0\) is
{%
\setlength{\belowdisplayskip}{-2ex} \setlength{\belowdisplayshortskip}{-2ex}
\setlength{\abovedisplayskip}{1ex} \setlength{\abovedisplayshortskip}{1ex}
\begin{align}
    y_i\left(t\right) = {\sum}^k_{j=1} x_j\left(t - \Delta_i\left(\phi_j\right)\right) + \eta_i\left(t\right), \text{ where }\label{eq:model}
\end{align}
}%
\begin{center}
  {\begingroup
    \setlength\tabcolsep{2pt} 
  \begin{tabular}{lp{0.7\linewidth}}
   \(\Delta_i\left(\phi_j\right) \in \mathbb{R}\) & is the inter-sensor delay for a signal impinging to the \(i\)th sensor from an angle of \(\phi_j \in \left[\nicefrac{-\rm\pi}{2}, \nicefrac{\rm\pi}{2}\right]\), \\
   \(x_j\left(t\right) \in \mathbb{R}\) & is the signal emitted by the \(j\)th source, and \\
   \(\eta_i\left(t\right) \in \mathbb{R}\) & is the noise and interference.
  \end{tabular}
  \endgroup}
\end{center}
In the DoA \textit{estimation} problem the source signals \( {\{ x_{j}\left(t\right) \}}_{j \in [k]} \), and noise and interference \( {\{ \eta_j\left(t\right) \}}_{j \in [M]}\) are all assumed to be unknown, and in the signal \textit{detection} problem, the number of sources, or model order, \(k\), is also unknown.
Note that the fact that we assume real-value signals means that we impose conjugate symmetry constraints~\citep[\S 2.8]{oppenheim_discretetime_2010} on the model.
This is without loss of generality, as our derivations can be generalized to complex signals by replacing all operations with their complex-valued counterparts and the Gaussians with circularly symmetric Gaussians.

\paragraph{Discrete Time--Frequency Model}\label{section:frequency_domain_model}
There are multiple ways to interpret the model in \cref{eq:model} in discrete time.
Previous works utilized the frequency domain representation of \cref{eq:model}.
That is, for each channel \(i \in [M]\), the received signal \(\vy_i\) is processed through the short-time Fourier transform (STFT) with non-overlapping ``snapshots'' such that \(y_i[b, s]\) is its \(b\)th frequency bin and \(s\)th snapshot.
For \(N = S \cdot B\) received samples, the STFT yields \(S\) snapshots and \(B\) frequency bins.
Then, for \( (s, b) \in [S] \times [B]\), the channel-wise concatenation \(\vy[s, b] \in \mathbb{C}^M\) is modeled as
{%
\setlength{\belowdisplayskip}{1.0ex} \setlength{\belowdisplayshortskip}{1.0ex}
\setlength{\abovedisplayskip}{1.0ex} \setlength{\abovedisplayshortskip}{1.0ex}
\begin{align}
    \vy\left[s, b\right] = {\textstyle \sum^k_{j=1}} \va\left(f_b, \phi_j\right) x_j\left[s,b\right] +
     \veta\left[s, b\right], \label{eq:complex_model}
\end{align}
}%
where \(\va\left(f_b, \phi_j\right) \in \mathbb{C}^M\) is a ``steering vector'' defined as 
{%
\setlength{\belowdisplayskip}{1.0ex} \setlength{\belowdisplayshortskip}{1.0ex}
\setlength{\abovedisplayskip}{1.0ex} \setlength{\abovedisplayshortskip}{1.0ex}
\begin{equation*}
  \va\left(f_b, \phi_j\right) 
  \triangleq 
  {\begin{bmatrix}
    \mathrm{e}^{ -\mathrm{i} 2 \mathrm{\pi} f_b \Delta_1\left(\phi_j\right) } & \ldots & \mathrm{e}^{ -\mathrm{i} 2 \mathrm{\pi} f_b \Delta_M\left(\phi_j\right) } 
  \end{bmatrix}}^{\top},
\end{equation*}
}%
where the upright $\mathrm{i} \triangleq \sqrt{-1}$ here denotes the imaginary unit.
For a uniform linear array (ULA), the inter-sensor delay is typically set as \(\Delta_i\left(\phi_j\right) \triangleq -\left(i - 1\right) d \sin\left(\phi_j\right) / c\), where \(c\) is the propagation speed of the medium (\textit{e.g.}, speed of sound in the case of acoustic signals) and \(d\) is the sensor displacement.

The model in \cref{eq:complex_model} has been widely used~\citep[\S 3.14.6.1]{chung_doa_2014}.
However, it has some limitations: When the delay \(\Delta_i\left(\phi_j\right) \cdot f_s\) is fractional, the ``steered'' signal is complex-valued, meaning that \cref{eq:complex_model} fails to model real-valued signals~\citep[\S 2.8]{oppenheim_discretetime_2010}.
Second, modeling the STFT forces us to reason about frequency resolution and temporal resolution, which should not inherently be necessary.
This work proposes an alternative, purely time-domain modeling approach that does not rely on the model of \cref{eq:complex_model}.

\vspace{-1ex}
\section{Bayesian Modeling}

\subsection{Time Domain Wideband Signal Model}\label{section:model}
\paragraph{Non-Causal Discrete-Time Model}
Recall the ideal continuous time model in \cref{eq:model}.
We will represent the time delay with an \(L\)-tap conjugate symmetric FIR filter, where \(L \in \mathbb{N}_{>0}\) is assumed to be an odd number for convenience.
Let the propagation of the signal emitted by the \(j\)th source to the \(i\)th sensor be represented as \(h_{i,j} \triangleq h_i\left(\phi_{j}\right)\), where \(i \in [M]\) and \(j \in [k]\).
\(h_{i,j}\) is assumed to be non-causally centered at \(n=0\) such that, if there is no delay, \(h_{i,j}[n] = \delta[n]\).
Then, for \(n \in \{0,\, \ldots,\, N-1\}\), the signal received by the \(i\)th sensor can be modeled as
{%
\setlength{\belowdisplayskip}{1ex} \setlength{\belowdisplayshortskip}{1ex}
\setlength{\abovedisplayskip}{1ex} \setlength{\abovedisplayshortskip}{1ex}
\begin{align}
    y_i\left[n\right] &= \sum^k_{j=1} \left( \sum_{l= -\lfloor{\nicefrac{L}{2}}\rfloor }^{ \lfloor{\nicefrac{L}{2}}\rfloor } h_{i,j}\left[l\right] x_j\left[n - l\right] \right) + \eta_i\left[n\right].
    \label{eq:noncausal_model}
\end{align}
}%
Notice that, to utilize all \(N\) measurements, the source signals \textit{have} to be defined for all \(n \in \{ - \lfloor \nicefrac{L}{2} \rfloor,\, \ldots ,\, N + \lfloor \nicefrac{L}{2} \rfloor \} \).

\paragraph{Periodic Discrete-Time Model}
Even though \cref{eq:noncausal_model} is non-causal, it is realizable by wrapping around, or ``aliasing,'' the non-causal part. 
This can be done by treating \(x_{j}\) and \( h_{i,j}\) as periodic signals with period \(N^{\prime} \triangleq N + L - 1\) such that
{%
\setlength{\belowdisplayskip}{1ex} \setlength{\belowdisplayshortskip}{1ex}
\setlength{\abovedisplayskip}{1ex} \setlength{\abovedisplayshortskip}{1ex}
\begin{align*}
    y_i\left[n\right] = \sum^k_{j=1} \left( \sum_{l=0}^{N^{\prime} - 1} h^{\prime}_{i,j} \left[l\right] x_j\left[ {\left(n - l\right)}_{\mathrm{mod} \; N^{\prime}}\right] \right) + \eta_i\left[n\right],
\end{align*}
}%
where \(h^{\prime}_{i,j} : [N^{\prime}] \to \mathbb{R}\) is \(h_{i,j}\) padded to the right with \(N\) zeros and the non-causal part wrapped around such that
{%
\setlength{\belowdisplayskip}{1.ex} \setlength{\belowdisplayshortskip}{1.ex}
\setlength{\abovedisplayskip}{1.ex} \setlength{\abovedisplayshortskip}{1.ex}
\begin{equation*}
    h^{\prime}_{i,j}[n] \triangleq \begin{cases}
        h_{i,j}[n] & \text{if \(0 \leq n \leq \lfloor \nicefrac{L}{2} \rfloor\)} \\
        0 & \text{if \( \lfloor \nicefrac{L}{2} \rfloor + 1 \leq n \leq N + \lfloor \nicefrac{L}{2} \rfloor - 1\)} \\
        h_{i,j}[n \text{-} N \text{-} L \text{+} 1] & \text{otherwise}.
    \end{cases}
\end{equation*}
}%
This corresponds to a circular convolution~\citep[\S 8.6.5]{oppenheim_discretetime_2010}:
{%
\setlength{\belowdisplayskip}{1ex} \setlength{\belowdisplayshortskip}{1ex}
\setlength{\abovedisplayskip}{1ex} \setlength{\abovedisplayshortskip}{1ex}
\begin{align}
    y_i\left[n\right] &= {\textstyle\sum^k_{j=1}} \left( h^{\prime}_{i,j} \circledast x_j \right) \left[n\right]  + \eta_i\left[n\right] \; .
    \label{eq:model_padded_circularconv}
\end{align}
}%
Note that, as we did not pad \(x_j\), this is \textit{not} the usual circular convolution realization of the linear convolution as in~\citep[\S 8.7]{oppenheim_discretetime_2010}.
Instead, we are assuming that the length of the source signal \(\vx\) is \textit{longer} than the received signal, depending on \(L\).
Compared to the more classical causal realization technique of introducing group delay~\citep[\S 5.5]{oppenheim_discretetime_2010}, this aliasing-based technique is computationally convenient under the discrete Fourier transform (DFT), as we will discuss in \cref{section:likelihood_computation}.

\paragraph{Linear System Representation}
Since \(h^{\prime}_{i,j}\) is a periodic FIR filter, it has a circulant matrix representation \(\mH_{i,j} \triangleq \mH_{i}\left(\phi_j\right) \in \mathbb{R}^{N^{\prime} \times N^{\prime}}\)~\citep{gray_toeplitz_2006}. 
Thus, we can write \cref{eq:model_padded_circularconv} as
{%
\setlength{\belowdisplayskip}{.5ex} \setlength{\belowdisplayshortskip}{.5ex}
\setlength{\abovedisplayskip}{.0ex} \setlength{\abovedisplayshortskip}{.0ex}
\begin{align*}
    \vy_{i} = {\sum_{j=1}^{k}} \mM \mH_{i,j} \,\vx_{j} + \veta \; ,
\end{align*}
}%
where \(\mM \triangleq {\begin{bmatrix} \boldupright{I}_{N} & \boldupright{O}_{L - 1} \end{bmatrix}} \in \mathbb{R}^{N \times N^{\prime}}\) is a ``truncation matrix'' that truncates the last \(L -1\) dimensions.

We can denote this as one big linear system:
{\small
\begin{align*}
    \begin{bmatrix}
        \vy_1 \\
        \vy_2 \\
        \vdots \\
        \vy_M
    \end{bmatrix} 
    =
    \begin{bmatrix}
        \mM \mH_{1,1} & \mM \mH_{1,2} &  \ldots & \mM \mH_{1,k} \\
        \mM \mH_{2,1} & \mM \mH_{2,2} &         & \mM \mH_{2,k} \\
        \vdots     &            & \ddots  & \vdots  \\
        \mM \mH_{M,1} & \mM \mH_{M,2} & \ldots  & \mM \mH_{M,k}
    \end{bmatrix}
    \begin{bmatrix}
        \vx_1 \\
        \vx_2 \\
        \vdots \\
        \vx_k
    \end{bmatrix}
    +
    \veta
    \; .
\end{align*}
}%
Denoting the block-wise truncation as \(\mM_{M} \triangleq \mathrm{block\text{-}diagonal}_M\left(\mM, \ldots, \mM\right)\) and the filters as \(\mH\left(\vphi_{1\text{:}k}\right) \triangleq \mH_{1\text{:}M,1\text{:}k}\), the resulting linear system is 
\begin{align}
    \vy = \mA\left(\vphi_{1\text{:}k}\right)\,\vx_{1\text{:}k} + \veta, \label{eq:big_system}
\end{align}
where \(\mA\left(\vphi_{1\text{:}k}\right) \triangleq \mM_{M} \mH\left(\vphi_{1\text{:}k}\right) : \mathbb{R}^{k} \rightarrow \mathbb{R}^{MN \times k N^{\prime} }\),
\(\vy \in \mathbb{R}^{MN}\), 
\(\vx_{1\text{:}k} \in \mathbb{R}^{k N^{\prime}}\).
The system in \cref{eq:big_system} is quite large, and attempts to solve the system directly will be computationally infeasible in scenarios of practical interest.
In \cref{section:likelihood_computation}, we will demonstrate how to speed up computation by leveraging the circulant structure of \(\mH_{i,j}\).

\paragraph{Time Delay Filter}
Recall that the time-delay filter \(h^{\prime}_{i,j}\) is an \(L\)-tap non-causal filter zero-padded into an \(N^{\prime}\)-tap periodic filter.
For our implementation, we directly used a \(N^{\prime}\)-tap realization of the periodic fractional delay filter by Pei and Lai~\citep{pei_closed_2014} (originally proposed in \citep{pei_closed_2012}; see also \citep{blok_comments_2013}) without padding.
Informally, it is a windowed-sinc filter:
\begin{align*}
  h_{i,j}[n] = w[n]\,\mathrm{sinc}\left(n - \Delta_i\left(\phi_j\right)\right) -  f_a\left(n\right),
\end{align*}
where \(\mathrm{sinc}\left(x\right) \triangleq \nicefrac{\sin\left(\mathrm{\pi}x\right)}{\left(\mathrm{\pi}x\right)}\), \(w[n]\) is some tapering window centered at \(n=0\), and \(f_a\) is a compensation term depending on \(a \in [0, 1]\) that trades pass-band ripple and transition sharpness.
As recommended by \citeauthor{pei_closed_2014}, we set \(a = 0.25\).
Also, we set \(L = N + 1\) for an even \(N \in \mathbb{N}_{>0}\) such that \(N^{\prime} = 2 N\).
Skipping padding is a suitable approximation as long as the sidelobes of the filter taper quickly enough.
Furthermore, this filter has a closed-form DFT~\citep[Eq. (6)]{pei_closed_2014}.

\vspace{-1.5ex}
\subsection{Hierarchical Bayesian Model}\label{section:bayesian_model} 
Based on the signal model in \cref{eq:big_system}, we present our Bayesian probabilistic model:
\vspace{-2ex}

\begin{center}
{%
\setlength{\belowdisplayskip}{.5ex} \setlength{\belowdisplayshortskip}{.5ex}
\setlength{\abovedisplayskip}{-.5ex} \setlength{\abovedisplayshortskip}{-.5ex}
\begin{alignat*}{3}
    &\lambda &&\sim \mathsf{Inv\text{-}Gamma}\left(\alpha_{\lambda}, \beta_{\lambda}\right)
    \\
    &k &&\sim \mathsf{Poisson}_{k \leq k_{\mathrm{max}}}\left(\lambda\right)
    \\
    &\sigma^2 &&\sim \mathsf{Inv\text{-}Gamma}\left( \alpha, {\beta} \right) &&
    \\
    &\phi_j  &&\sim \mathsf{Uniform}\left[-\frac{\pi}{2}, \frac{\pi}{2} \right]  &&\; \text{for}\; j = 1, \ldots, k
    \\
    &\gamma_j  &&\sim p\left(\gamma\right) &&\; \text{for}\; j = 1, \ldots, k
    \\
    &\vx_{1\text{:}k} &&\sim \mathcal{N}\left( \mathbf{0}_{k (N+L-1)}, \sigma^2 \mSigma\left(\vgamma_{1\text{:}k}\right) \right)  &&
    \\
    &\vy  \;&&\sim\; \mathcal{N}\left( \mA\left( {{\vphi}_{1\text{:}k}} \right) \, {\vx_{1\text{:}k}}, \sigma^2 \boldupright{I}_{MN} \right),  &&
\end{alignat*}
}%
\end{center}

%
\noindent%
where \(\gamma_j\) is a parameter local to the \(j\)th source, \(\alpha\), \(\beta\), \(\alpha_{\lambda}\), and \(\beta_{\lambda}\) are hyperparameters.
Notice that imposing an inverse-gamma prior on $\sigma^2$ implies that the marginal priors of the noise and source signals $x_{1:k}$ correspond to Student-$t$ distributions.
Furthermore, assumptions on the spectral structure of the latent source signals can be expressed by manipulating the covariance matrix \(\mSigma\left(\vgamma_{1\text{:}k}\right)\).
We will later discuss our choice of \(\mSigma\left(\vgamma_{1\text{:}k}\right)\), its hyperparameters \(\gamma_1,\, \ldots,\, \gamma_k\), and the corresponding hyperprior \(p\left(\gamma\right)\).
Since we primarily care about \(k\) and \(\phi_1, \ldots, \phi_k\), the rest are considered nuisance variables.

\vspace{0.8ex}
\paragraph{Prior on Model Order}
For the number of sources, or \textit{model order}, \(k\), we follow~\citep{andrieu_joint_1999,roodaki_joint_2010} and assign a truncated Poisson-Gamma mixture prior.
Due to conjugacy, we can marginalize out \(\lambda\), which yields:
{%
\setlength{\belowdisplayskip}{1ex} \setlength{\belowdisplayshortskip}{1ex}
\setlength{\abovedisplayskip}{1ex} \setlength{\abovedisplayshortskip}{1ex}
\begin{equation*}
     k \sim \mathsf{NB}_{k \leq k_{\mathrm{max}}}\big(\alpha_{\lambda} , {\beta_{\lambda}}/{\left(\beta_{\lambda} + 1\right)}\big),
\end{equation*}
}%
where \(\mathsf{NB}_{k \leq k_{\mathrm{max}}}\left(r, p\right)\) is a negative binomial distribution truncated at \(k_{\mathrm{max}}\) with \(r > 0\) successes until stopping with a success probability \(p \in [0, 1]\).
Following~\citep{andrieu_joint_1999}, we set \(\alpha_{\lambda} = \nicefrac{1}{2} + 0.1\) and \(\beta_{\lambda} = 0.1\).
This choice of prior is motivated by the following reasoning: We assume that it is always more likely to encounter fewer targets than encountering more targets.
This aligns with the null hypothesis choice in hypothesis test-based detection approaches~\citep{viberg_detection_1991,ottersten_exact_1993,boehme_statistical_1995,maiwald_multiple_1994,chung_detection_2007,shumway_replicated_1983}.
Therefore, any prior with monotonically decreasing probability mass is admissible.
Among these, the negative binomial with \(r < 1\) exhibits a heavier tail than alternatives such as the geometric distribution. 

\paragraph{Prior on the Source Signals}\label{section:source_signal_prior}

For the prior on the source signals \(\vx_{1\text{:}k}\),~\citep{andrieu_joint_1999,ng_wideband_2005} have used the Zellner's \(g\)-prior~\citep{zellner_assessing_1986,nielsen_bayesian_2014}.
When using Bayes factors (or RJMCMC), the \(g\)-prior is appealing as it has been shown to have good frequentist properties such as consistency of model selection~\citep{sparks_necessary_2015,liang_mixtures_2008}.
Unfortunately, the classic \(g\)-prior assumes that the linear system is full-rank, which is not the case for \(\mA\left(\vphi_{1\text{:}k}\right)\); the blocks of \(\mA\left(\vphi_{1\text{:}k}\right)\) are not full-rank since the truncation matrix \(\mM\) is rank-deficient, and the DC components of \(\vx_{1\text{:}k}\) are unidentifiable under \(\mH_{i,j}\).
Instead, we set the source signal covariance as
%
%
\begingroup
{
\setlength\arraycolsep{2pt}
{%
\setlength{\belowdisplayskip}{1ex} \setlength{\belowdisplayshortskip}{1ex}
\setlength{\abovedisplayskip}{1ex} \setlength{\abovedisplayshortskip}{1ex}
\begin{align*}
    \mSigma\left(\vgamma_{1\text{:}k}\right)
    &\triangleq
    \text{block-diagonal}_k\left( \gamma_1 \mathbf{I}_{N^{\prime}},\; \ldots,\; \gamma_k \mathbf{I}_{N^{\prime}} \right) \; ,
\end{align*}
}%
}%
where the hyperparameters \( \vgamma_{1\text{:}k} = \left(\gamma_1, \ldots, \gamma_k\right) \) are independently sampled from some hyperprior \(p\left(\gamma_j\right)\).
Our choice of \(\mSigma\left(\vgamma_{1\text{:}k}\right)\) assumes that the source signals independently follow a Gaussian process with a white-noise-like spectral structure.
Our experiments in \cref{section:detection_narrowband} demonstrate that this choice works well even for narrowband signals.
Also, since \(\mSigma\) is multiplied with \(\sigma\), the noise power, the signal coming from the \(j\)th source has a power of \(\sigma^2 \gamma_j\).
Therefore, similarly to the \(g\) parameter in \(g\)-prior-based signal detection models~\citep{nielsen_bayesian_2013,nielsen_bayesian_2014}, \(\gamma_j\) can be directly interpreted as the SNR of the \(j\)th source.

The hyperprior \(p(\gamma_j)\) is known to greatly affect the performance of Bayesian model comparison~\citep[\S 5]{kass_bayes_1995}.
In our case, the SNR interpretation of \(\gamma_j\) provides a straightforward way to set the hyperprior.
Recall that, under a decision-theoretic perspective, any Bayesian point estimator obtained from the posterior of our model will achieve the best average performance over the range of SNRs supported by \(p(\gamma_j)\).
Therefore, it suffices to choose \(p(\gamma_j)\) such that it has a high probability over the effective operational range (in SNR) of the system.
Here, we evaluate the non-informative choice of
{%
\setlength{\belowdisplayskip}{1ex} \setlength{\belowdisplayshortskip}{1.ex}
\setlength{\abovedisplayskip}{1ex} \setlength{\abovedisplayshortskip}{1ex}
\begin{equation*}
    p\left(\gamma_j\right) = \mathsf{Inv\text{-}Gamma}\left(\gamma_j;\, \alpha_{\gamma}, \beta_{\gamma}\right) \, ,
\end{equation*}
}%
with \(\alpha_{\gamma} = \beta_{\gamma} = \epsilon\) for a small \(\epsilon\), in our case, \(\epsilon = 10^{-2}\).
This mimics Jeffrey's prior $p(\gamma_j) \propto 1/\gamma_j$~\citep{consonni_prior_2018}, which is uninformative in the sense that it is invariant to the scaling of $\gamma_j$.
As such, our choice of prior should result in reasonable performance across a wide range of SNRs.
By contrast, informative priors may improve performance in specific ranges of SNRs by trading off performance outside of those regions.
We evaluate our choice of prior in \cref{section:snr_priors}.

We remark that our choice of source priors is preliminary, and there are multiple immediate directions for improvement.
For instance, \citep{maruyama_fully_2011,baragatti_study_2012,gupta_information_2009,som_paradoxes_2014} developed various extensions to the classic \(g\)-prior that do not assume full-rankness of \(\mA\left(\vphi_{1\text{:}k}\right)\).
Also, complex spectral structures can be expressed via Gaussian process priors~\citep{wilson_gaussian_2013,remes_nonstationary_2017,tobar_bandlimited_2019}.
Investigating these priors would be an interesting future direction.

\vspace{-2ex}
\subsection{Marginalization of Nuisance Variables}\label{section:marginalization}
The nuisance variables \(\sigma^2\) (noise variance) and \(\vx_{1\text{:}k}\) (source signals) in our model can be marginalized away.
The marginalization process is crucial for the computational and statistical efficiency of the process; the number of parameters entering the likelihood is reduced from \(k \left(N + L - 1\right) + 2 k + 1\) to \(2 k\).
This process is also known as \textit{collapsing} or \textit{Rao-Blackwellization}~\citep[\S 9.3]{robert_monte_2004}.

\paragraph{Conjugate Analysis}
Once the system \(\mA \triangleq \mA\left(\vphi_{1\text{:}k}\right)\) has been formed, the model is equivalent to a Bayesian linear regression model conditioned on \(\vphi_{1\text{:}k}\), \(\vgamma_{1\text{:}k}\) such that the conditional joint likelihood is
{%
\setlength{\belowdisplayskip}{1ex} \setlength{\belowdisplayshortskip}{1ex}
\setlength{\abovedisplayskip}{1ex} \setlength{\abovedisplayshortskip}{1ex}
\begin{align*}
    &p\left(\vy,\, \vx_{1\text{:}k},\, \sigma^2 \mid \phi_{1\text{:}k},\, \vgamma_{1\text{:}k},\, \alpha,\, \beta \right)
    \\
    &=
    p\left(\vy \mid \vx_{1\text{:}k}, \phi_{1\text{:}k}, \sigma^2 \right)
    p\left(\vx_{1\text{:}k} \mid \sigma^2, \vgamma_{1\text{:}k} \right)
    p\left(\sigma^2 \mid \alpha, \beta\right)
    \\
    &=
    \mathcal{N}\left(\vy \mid \mA\left(\phi_{1\text{:}k}\right) \vx_{1\text{:}k}, \sigma^2 \boldupright{I}_{MN} \right)
    \\
    &\times 
    \mathcal{N}\left(\vx_{1\text{:}k} \mid \boldupright{0}_{k \left(N+L-1\right)},\,  \sigma^2\mSigma\left(\vgamma_{1\text{:}k}\right)\right)
    \mathsf{Inv\text{-}Gamma}\left(\sigma^2; \alpha, \beta\right) .
\end{align*}
}%
Through conjugate analysis, the joint conditional posterior of \(\vx_{1\text{:}k}\) and \(\sigma^2\) is found to be a normal-inverse-gamma joint posterior~\citep[Eq. (7.69)-(7.73)]{murphy_machine_2012}
{%
\setlength{\belowdisplayskip}{1ex} \setlength{\belowdisplayshortskip}{1ex}
\setlength{\abovedisplayskip}{1ex} \setlength{\abovedisplayshortskip}{1ex}
\begin{align}
  &p\left( \vx_{1\text{:}k}, \sigma^2 \mid \vy, k, \vphi_{1\text{:}k}, \vgamma_{1\text{:}k}, \alpha, \beta  \right)
  \nonumber
  \\
  &\qquad\qquad=
  \mathsf{NIG}\left(
    \vx_{1\text{:}k}, \sigma^2 \mid \widetilde{\vmu}, \widetilde{\mSigma}, \widetilde{\alpha}, \widetilde{\beta}
  \right),
  \label{eq:conditionalposterior}
\end{align}
}%
where the parameters are given as
{%
\setlength{\belowdisplayskip}{.0ex} \setlength{\belowdisplayshortskip}{.0ex}
\setlength{\abovedisplayskip}{.5ex} \setlength{\abovedisplayshortskip}{.5ex}
\begin{alignat*}{3}
  \widetilde{\vmu} &\triangleq \widetilde{\mSigma} \, \mA^{\herm} \vy,
  &&\widetilde{\alpha} &&\triangleq \alpha + \frac{N \, M}{2},
  \\
  \widetilde{\mSigma} &\triangleq {\left({\mSigma\left(\vgamma_{1\text{:}k}\right)}^{-1} + \mA^{\herm} \mA \right)}^{-1},\;\quad
  &&\widetilde{\beta}  &&\triangleq \beta + \frac{1}{2} \vy^{\herm} \mP_{\bot} \vy,
\end{alignat*}
}%
{%
\setlength{\belowdisplayskip}{1ex} \setlength{\belowdisplayshortskip}{1ex}
\setlength{\abovedisplayskip}{.0ex} \setlength{\abovedisplayshortskip}{.0ex}
\begin{align}
  \mP_{\bot} 
  \triangleq
  {\big( \mA^{\herm} \mSigma \mA + \boldupright{I} \big)}^{-1}
  =
  \boldupright{I} - \mA {\left( {\mSigma}^{-1} + \mA^{\herm} \mA \right)}^{-1} \mA^{\herm}.
  \label{eq:pbot}
\end{align}
}%
Here, \(\mP_{\bot}\) can be interpreted as a regularized orthogonal complement of the column-space projection of \(\mA\).

\paragraph{Collapsed Likelihood}
From the joint posterior \cref{eq:conditionalposterior}, it is possible to marginalize away both \(\vx_{1\text{:}k}\) and \(\sigma^2\), resulting in the unnormalized \textit{collapsed likelihood}.
This is given by the normalizing constants in \cref{eq:conditionalposterior},
{%
\setlength{\belowdisplayskip}{1ex} \setlength{\belowdisplayshortskip}{1ex}
\setlength{\abovedisplayskip}{1ex} \setlength{\abovedisplayshortskip}{1ex}
\begin{align}
  &p\left(\vy \mid k, \vphi_{1\text{:}k}, \vgamma_{1\text{:}k}, \alpha, \beta \right) \nonumber \\ 
  &\qquad\qquad\propto
  {
    \mathrm{det}\left(
      \mP_{\bot}
    \right)
  }^{-\nicefrac{1}{2}}
  {\left(
    \nicefrac{\beta}{2}
    +
    \vy^{\herm}
    \mP_{\bot}
    \vy
  \right)}^{- M N / 2 + \alpha} \, ,
  \label{eq:collapsed_likelihood}
\end{align}
}%
which is reminiscent of the posterior of Andrieu and Doucet~\citep{andrieu_joint_1999}.
Given some order \(k\), this now reduces the number of parameters that need to be inferred from \(k (N + L - 1) + 2 k + 1\) to \(2 k\).
Intuitively, the collapsed likelihood measures the signal power that is reduced by the regularized system.
Since we are still left to deal with a matrix inverse and determinants, in \cref{section:likelihood_computation}, we will present a way to efficiently compute this collapsed likelihood.

\paragraph{Conditional Posterior of the Source Signals}
The latent source signals can readily be reconstructed by sampling from the conditional posterior of \(\vx_{1\text{:}k}\).
From the properties of the normal-inverse-gamma, the collapsed conditional posterior for \(\vx_{1\text{:}k}\) is a Student-\(t\) distributions given as
{%
\setlength{\belowdisplayskip}{1ex} \setlength{\belowdisplayshortskip}{1ex}
\setlength{\abovedisplayskip}{1ex} \setlength{\abovedisplayshortskip}{1ex}
\begin{align}
  &\pi\left( \vx_{1\text{:}k} \mid \vy, k, \vphi_{1\text{:}k}, \vgamma_{1\text{:}k}, \alpha, \beta  \right) 
  \nonumber
  \\
  &\qquad\qquad=
  {\mathsf{Student\text{-}}t}_{2 \widetilde{\alpha}}\big(\vx_{1\text{:}k};
   \widetilde{\mu}_{\vx}, \big( {\widetilde{\beta}} /
    {\widetilde{\alpha}}\big) \widetilde{\mSigma}_{\vx}
  \big),
  \label{eq:reconstruction_posterior}
\end{align}
}%
where \({\mathsf{Student\text{-}}t}_{\nu}\left(\cdot; \vm, \mV\right)\) is the density of a Student-\(t\) distribution with \(\nu\)-degrees of freedom with location \(\vm\) and scale \(\mV\)~\citep[Eq. (7.75)]{murphy_machine_2012}.
We detail the reconstruction procedure in \cref{section:detection_estimation_procedure}.


\vspace{-1ex}
\section{Bayesian Computation}\label{section:computation}
The probabilistic model described in \cref{section:bayesian_model} forms a posterior distribution
{%
\setlength{\belowdisplayskip}{1ex} \setlength{\belowdisplayshortskip}{1ex}
\setlength{\abovedisplayskip}{1ex} \setlength{\abovedisplayshortskip}{1ex}
\begin{align}
    &\pi\left(
        k, \vphi_{1\text{:}k}, \vgamma_{1\text{:}k} \mid \vy, \alpha, \beta, \alpha_{\lambda}, \beta_{\lambda}
    \right)
    \nonumber
    \;
    \propto 
    \;
    p\left(\vy \mid k, \vphi_{1\text{:}k}, \vgamma_{1\text{:}k}, \alpha, \beta \right)
    \nonumber
    \\
    &\quad\times
    p\left(k; \alpha_{\lambda}, \beta_{\lambda}\right)
    {\textstyle\prod^{k}_{j=1}} p\left(\gamma_j; \alpha_{\gamma}, \beta_{\gamma}\right) p\left(\phi_j\right).
    \label{eq:posterior}
\end{align}
}%
Our detection procedure will follow from this posterior, which is intractable. 
Therefore, we will replace it with its Monte Carlo approximation over a collection of samples, \({\{ (k^t, \vphi_{1\text{:}k^t}^t, \vgamma_{1\text{:}k^t}^t) \}}_{t \geq 0}\), drawn via RJMCMC.
Overall, in this section, we will discuss the following:
\begin{enumerate}
    \item \cref{section:detection_estimation_procedure}: Detection, estimation, and reconstruction using samples from the posterior in \cref{eq:posterior}.
    \item \cref{section:likelihood_computation}: Efficiently computing the collapsed likelihood \cref{eq:collapsed_likelihood}.
    \item \cref{section:rjmcmc}: Drawing samples from the posterior in \cref{eq:posterior} using RJMCMC.
\vspace{-1ex}
\end{enumerate}

\vspace{-1ex}
\subsection{Detection, Estimation, and Reconstruction Procedures}\label{section:detection_estimation_procedure}
\paragraph{Detection}
We view the detection problem as estimating the number of sources.
From a decision-theoretic perspective, this can be formulated as a decision problem
{%
\setlength{\belowdisplayskip}{.5ex} \setlength{\belowdisplayshortskip}{.5ex}
\setlength{\abovedisplayskip}{.5ex} \setlength{\abovedisplayshortskip}{.5ex}
\begin{align}
    \widehat{k} \triangleq \argmin_{j \in \left\{0,\; \ldots,\; k_{\mathrm{max}}\right\}} {\sum^{k_{\mathrm{max}}}_{k=0}} l\left(k, j \right) \pi\left(k \mid \vy, \alpha, \beta\right)
    \label{eq:decision_problem}
\end{align}
}%
given the loss function \(l : \mathbb{N}_{\geq 0} \times \mathbb{N}_{\geq 0} \to \mathbb{R}_{>0}\) and the marginal posterior \(\pi\left(k \mid \vy, \alpha, \beta, \alpha_{\lambda}, \beta_{\lambda}\right)\).
Then choosing \(\widehat{k}\) as the posterior mode corresponds setting \(l\) as the 0--1 loss \(l\left(k, j\right) = \mathds{1}_{j \neq k}\), while the posterior median corresponds to the robust \(\ell_1\) loss \(l\left(k, j\right) = \abs{k -j}\)~\citep[\S 2.5]{robert_bayesian_2001}.
The solution to this decision problem is \textit{Bayes optimal}: it achieves the lowest average loss when the average is taken over the prior and all random data realizations.
For our experiments, we use the \(\ell_1\) loss, which means we choose \(k^{\prime}\) to be the posterior median, also known as the ``median probability model'' \citep{barbieri_optimal_2004,ghosh_bayesian_2015}.
Naturally, the expectation in \cref{eq:decision_problem} is intractable, and is therefore Monte Carlo-approximated using the RJMCMC samples.

\begin{figure}[t]
    \vspace{-4ex}
    \centering
    \includegraphics[scale=1.0]{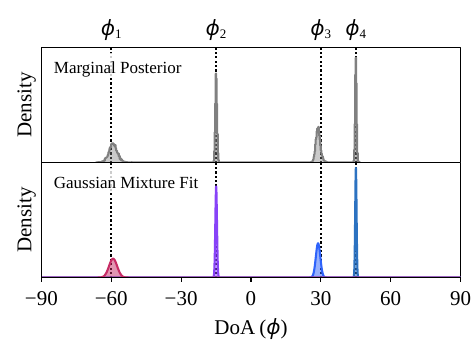}
    \vspace{-1.5ex}
    \caption{\textbf{Demonstration of RJMCMC output.}
    (\textbf{top}) Histogram of the posterior samples of the DoAs and (\textbf{bottom}) its Gaussian mixture approximation after relabeling.
    The data was generated from 4 sources with \(\phi_j \in \{-60^{\circ}, -15^{\circ}, 30^{\circ}, 45^{\circ}\}\) with a corresponding SNR of \(\gamma_j \in \{-4 \text{ dB}, 0 \text{ dB}, -6 \text{ dB}, 4 \text{ dB}\}\). 
    }
    \label{figure:output_doa_density}
    \vspace{-3ex}
\end{figure}

\paragraph{Estimation}
Although estimation is not the main interest of this article, it can also be performed by inspecting the marginal density of the DoAs, for any $k \geq 0$, \(\phi_1, \ldots, \phi_k\), or the DoAs conditional on the output of the decision problem \(\phi_1, \ldots, \phi_{\widehat{k}} \mid \widehat{k}\).
The former corresponds to performing Bayesian model averaging, while the latter corresponds to Bayesian model selection.
This can be done using the same RJMCMC samples used for the detection problem.
One issue when analyzing the DoAs through RJMCMC samples is that the individual DoA \(\phi_j\) is subject to ``label switching.''
That is, for each posterior sample \(\vtheta_k = (\vphi_{1\text{:}k}, \vgamma_{1\text{:}k})\), there is no way to know to which source the DoAs \(\phi_1, \ldots, \phi_k\) belong to.
Therefore, to analyze the parameters of a specific \textit{source}, it is necessary to relabel each DoA \(\phi_{j}\) and its local variable \(\gamma_{j}\).

For relabeling, one can use the mixture model fitting procedure by Roodaki \textit{et al.}~\citep{roodaki_relabeling_2014}.
This procedure assumes that the \(k\) DoAs \(\vphi_{1\text{:}k} = (\phi_{1}, \ldots, \phi_{k})\) in \(\vtheta_k\) are sampled without replacement from a mixture of \(\kappa \in \mathbb{N}_{>0}\) Gaussian components and a uniform component representing outliers.
For \(\kappa\), we choose the 90\% percentile of the posterior of \(k\), as proposed in~\citep{roodaki_relabeling_2014}. 
An example is demonstrated in \cref{figure:output_doa_density}: In contrast to the ``unlabeled'' posterior (gray), the components of the mixture approximation (colored) can be used to quantify the uncertainty or obtain point estimates of the individual sources.

\paragraph{Reconstruction}
The latent source signals can be reconstructed through the following sampling procedure:
{%
\setlength{\belowdisplayskip}{1ex} \setlength{\belowdisplayshortskip}{1ex}
\setlength{\abovedisplayskip}{1ex} \setlength{\abovedisplayshortskip}{.5ex}
\begin{alignat*}{2}
     k, \vphi_{1\text{:}k}, \vgamma_{1\text{:}k} &\sim \pi\left(k, \vphi_{1\text{:}k}, \vgamma_{1\text{:}k} \mid \vy, \alpha, \beta, \alpha_{\lambda}, \beta_{\lambda}\right)
     \\
     \vx_{1\text{:}k} &\sim \pi\left(\vx_{1\text{:}k} \mid \vy, k, \vphi_{1\text{:}k}, \vgamma_{1\text{:}k}, \alpha, \beta \right).
\end{alignat*}
}%
That is, for each RJMCMC sample, the closed-form conditional in \cref{eq:reconstruction_posterior} can be used to sample from the latent source signal posterior.
The ``labels'' generated by the relabeling procedure in \citep{roodaki_relabeling_2014} can be used to register the source signal samples to their corresponding source.

\vspace{-1.5ex}
\subsection{Computing the Likelihood in the Frequency Domain}\label{section:likelihood_computation}

The collapsed likelihood in~\cref{eq:collapsed_likelihood} involves matrix operations that are quite costly.
For instance, naively computing \(\mP_{\bot}\) has a complexity of \(\mathcal{O}(N^3 k^3)\).
We will now discuss how to improve this by leveraging the frequency domain representation of FIR filters.

\paragraph{Stripe Decomposition}
The matrix representation of an FIR filter with periodic coefficients exhibits a structure known as a circulant matrix~\citep[\S 3]{gray_toeplitz_2006}.
Furthermore, a circulant matrix such as \(\mH_{i,j}\) admits a spectral decomposition
{%
\setlength{\belowdisplayskip}{.5ex} \setlength{\belowdisplayshortskip}{.5ex}
\setlength{\abovedisplayskip}{.5ex} \setlength{\abovedisplayshortskip}{.5ex}
\begin{equation*}
    \mH_{i,j} = \mW \mS_{i,j} \mW^{\herm},
\end{equation*}
}%
where \(\mS_{i,j}\) is the spectrum of \(h^{\prime}_{i,j}\), and \(\mW^{\herm}, \mW \in \mathbb{C}^{N^{\prime} \times N^{\prime}}\) are the \textit{normalized} \(N^{\prime}\)-point DFT matrix and its inverse.
Recall that our choice of filter has a closed form DFT provided in \citep{pei_closed_2014}.
Therefore, we can directly compute \(\mS_{i,j}\) from \(\phi_j\).
From these facts, the concatenation \(\mH \triangleq \mH_{1\text{:}M, 1\text{:}k}\) admits a matrix decomposition we call a \textit{stripe decomposition} through 
\cref{thm:eigendecomp_to_stripematrix} in \cref{section:appendix_stripe}.
That is, there exists a matrix \(\mS \in \mathbb{C}^{M N^{\prime} \times k N^{\prime}}\) such that
{%
\setlength{\belowdisplayskip}{.5ex} \setlength{\belowdisplayshortskip}{.5ex}
\setlength{\abovedisplayskip}{0ex} \setlength{\abovedisplayshortskip}{0ex}
\begin{align}
    \mH = \mPhi_{M}\,\mS\,\mPhi_{k}^{\herm}, 
    \label{eq:stripe_decomposition}
\end{align}
}%
where \(\mPhi_M^{\herm} \in \mathbb{C}^{M N^{\prime} \times M N^{\prime}}\) is a block-wise DFT defined as
{%
\setlength{\belowdisplayskip}{.5ex} \setlength{\belowdisplayshortskip}{.5ex}
\setlength{\abovedisplayskip}{.5ex} \setlength{\abovedisplayshortskip}{.5ex}
\begin{align*}
    \mPhi_{M}^{\herm} \triangleq \mathrm{block\text{-}diagonal}_{M} \left( \mW^{\herm}, \ldots, \mW^{\herm} \right),
\end{align*}
}%
is a unitary matrix, \(\mS\) is a stripe matrix such that each \(i,j\)th block for \((i,j) \in [M] \times [k]\) is a diagonal matrix
{%
\setlength{\belowdisplayskip}{.5ex} \setlength{\belowdisplayshortskip}{.5ex}
\setlength{\abovedisplayskip}{.5ex} \setlength{\abovedisplayshortskip}{.5ex}
\[
    {[\mS]}_{i,j} \triangleq \mS_{i,j} = \mathrm{diagonal}\left(s_{i,j}[0], \ldots, s_{i,j}[N^{\prime}\text{-}1]\right),
\]
}%
where \(s_{i,j}[0], \ldots, s_{i,j}[N^{\prime}-1]\) are the DFT coefficients of \(h_{i,j}^{\prime}\).
Again, for our choice of time delay filter (\cref{section:model}), these have a closed form given \(\phi_j\)~\citep{pei_closed_2014}.
Crucially, \(\mS\) is not a diagonal matrix nor a block diagonal matrix, but a sparse structure referred to as (diagonally) striped by~\citet{smolarski_diagonallystriped_2006}.
We will refer to such structured matrix as a \textit{stripe matrix}.

Similarly, our source signal covariance \(\mSigma \triangleq \mSigma\left(\vgamma_{1\text{:}k}\right)\) also admits a stripe decomposition 
{%
\setlength{\belowdisplayskip}{.5ex} \setlength{\belowdisplayshortskip}{.5ex}
\setlength{\abovedisplayskip}{.0ex} \setlength{\abovedisplayshortskip}{.0ex}
\begin{equation*}
    \mSigma = \mPhi_k \mT \, \mPhi_k^{\herm},
\end{equation*}
}%
where the stripe matrix \(\mT \in \mathbb{R}^{k N^{\prime} \times k N^{\prime}}\) is symmetric block-diagonal where each \(j\)th block for \(j \in [k]\) is given as
{%
\setlength{\belowdisplayskip}{.5ex} \setlength{\belowdisplayshortskip}{.5ex}
\setlength{\abovedisplayskip}{.5ex} \setlength{\abovedisplayshortskip}{.5ex}
\begin{equation*}
    {[\mT]}_{j,j} = \gamma_j \, \boldupright{I}_{N^{\prime}}.
\end{equation*}
}%
We also point out that \(\mT\) is positive definite (PD).

\begin{figure}
    \vspace{-1ex}
    \centering
    \scalebox{1.0}{
    \begin{tikzpicture}
    \begin{groupplot}[
        group style={
            group size=1 by 3,
            vertical sep=4ex,
            xlabels at=edge bottom,
            ylabels at=edge left,
        },
        xmin  = 0,
        xmax  = 63,
        ymin  = -3,
        ymax  = 3,
        width = 0.9\columnwidth,
        height= 28ex,
        xlabel={Sample Index \(n\)},
        ylabel={Amplitude},
        xtick pos=bottom,
        ytick pos=left,
        major tick length=2pt,
        xtick ={0,15,31,47,63},
        ytick ={-3,-2,-1,0,1,2,3},
        xtick align=outside,
        ytick align=outside,
        axis line style = thick,
        every tick/.style={black,thick},
        legend columns=0,
        legend style={
            text opacity=1.0,
            draw=none,
            anchor=center,
            at={(0.5,1.2)},
        },
    ]
    
    

    
    \nextgroupplot[]
    \addplot[
        very thick,
        cherry1,
    ] table [
        x expr=\thisrowno{0}, 
        y expr=\thisrowno{1}, 
        col sep=comma,
    ] {figures/signal_reconstruction_snr=-4.csv};
    
    \addplot[
        mark size=1.5pt,
        very thick,
        opacity=0.8,
        cherry3,
    ] table [
        x expr=\thisrowno{0}, 
        y expr=\thisrowno{2}, 
        col sep=comma,
    ] {figures/signal_reconstruction_snr=-4.csv};
    
    \legend{True, MMSE}
    
    \foreach \n in {3, ..., 100} 
    {
    \addplot[
        cherry3,
        opacity=0.05,
    ] table [
        x expr=\thisrowno{0}, 
        y expr=\thisrowno{\n}, 
        col sep=comma,
    ] {figures/signal_reconstruction_snr=-4.csv};
    }

    \draw[-Latex] (rel axis cs: 0.70,0.85)--(rel axis cs: 0.85,0.8);
    \node[opacity=0.8, text opacity=1.0, fill=white, anchor=east] at (rel axis cs: 0.70,0.85) {\small{}approximation error};
    
    
    
    
\end{groupplot}
\end{tikzpicture}
    }
    \caption{
    \textbf{Reconstructions of a Latent Source Signal.}
    The data was simulated from a single wideband signal at \(\phi = -\pi/4\) with bandwidth [10 Hz, 1 kHz] and an SNR of $-4$ dB and approximated the resulting posterior with a chain of \(10^3\) MCMC samples after \(10^3\) burn-in steps.
    We then computed the source signal posterior according to the procedure in \cref{section:detection_estimation_procedure}.
    MMSE, the minimum mean-square error estimate, is the posterior mean, while the lightly shaded lines are 100 random samples from the reconstruction posterior.
    }
    \label{fig:reconstruction}
    \vspace{-3ex}
\end{figure}

Stripe-\textit{decomposable} matrices can be transposed, added, and multiplied efficiently in the space of stripe matrices.
Furthermore, their inverse and determinant also directly follow from the inverse and determinant of the corresponding stripe matrix.
(See \cref{thm:stripe_properties} in \cref{section:appendix_stripe}).

\paragraph{Approximate Collapsed Likelihood}
We can now represent the full system matrix using stripe matrices as
{%
\setlength{\belowdisplayskip}{.5ex} \setlength{\belowdisplayshortskip}{.5ex}
\setlength{\abovedisplayskip}{.5ex} \setlength{\abovedisplayshortskip}{.5ex}
\begin{align}
    \mA = \mM_{M} \, \mPhi_{M} \mS \, \mPhi_{k}^{\herm} \; .
    \label{eq:system_stripe_identity}
\end{align}
}%
Unfortunately, the truncation operation \(\mM_M\) complicates computation as it does not admit a stripe decomposition.
This is because truncation cannot be represented as an LTI system.
Therefore, we further rely on the approximation
{%
\setlength{\belowdisplayskip}{0ex} \setlength{\belowdisplayshortskip}{0ex}
\setlength{\abovedisplayskip}{1ex} \setlength{\abovedisplayshortskip}{1ex}
\begin{align}
    \mM_{M}^{\top} \mM_{M} \approx \boldupright{I}_{M N^{\prime}}.
    \label{eq:non_truncation_approximation}
\end{align}
}%
\vspace{-1.5ex}
\begin{remark}\label{remark:mean_reverting}
    The basic consequence of the approximation in \cref{eq:non_truncation_approximation} is that the system does not ignore the zero padding in \(\vy_{\mathcal{F}}\).
    Therefore, the non-causal part of the latent source signal will revert towards zero with high confidence.
\end{remark}

The effect noted in \cref{remark:mean_reverting} is visualized in 
\cref{fig:reconstruction}.
Here, we reconstructed a source signal according to the procedure in   \cref{section:detection_estimation_procedure}.
We can see that the reconstructed signal becomes inaccurate near the end due to a zero-reverting behavior.
In low-SNR situations, however, the uncertainty in the posterior will dominate the modeling error.
Therefore, the modeling error should only be a problem at high SNRs.

\vspace{1ex}
Now, the collapsed likelihood is approximately given as
{%
\setlength{\belowdisplayskip}{.5ex} \setlength{\belowdisplayshortskip}{.5ex}
\setlength{\abovedisplayskip}{.5ex} \setlength{\abovedisplayshortskip}{.5ex}
\begin{align}
    &\text{\cref{eq:collapsed_likelihood}}
    \approx
    {\left( \mathrm{det}\left( \mT^{-1} \right)/\mathrm{det}\left( \mT^{-1} + \mS^{\herm} \, \mS \right) \right)}^{\nicefrac{1}{2}} 
    \nonumber
    \\
    &\;\times
    {\left(\frac{\alpha}{2} + \vy^{\herm} \vy
    -
    \vy_{\mathcal{F}}^{\herm} \, \mS \,
    {\left( \mT^{-1}  + \, \mS^{\herm} \mS \right)}^{-1}
     \mS^{\herm} \vy_{\mathcal{F}} \right)}^{-\nicefrac{MN}{2} + \beta},
     \label{eq:approximate_collapsed_likelihood}
\end{align}
}%
where \(\vy_{\mathcal{F}} \triangleq \mPhi_{M}^{\herm} \mM^{\top}_{M} \vy \in \mathbb{C}^{M N^{\prime}}\)  is the padded blockwise-normalized-DFT of \(\vy\).
The full derivation is in \cref{section:approximate_collapsed_likelihood}.

\paragraph{Computation with the LDL\(^{\top}\) Decomposition}
Let 
{%
\setlength{\belowdisplayskip}{.5ex} \setlength{\belowdisplayshortskip}{.5ex}
\setlength{\abovedisplayskip}{.5ex} \setlength{\abovedisplayshortskip}{.5ex}
\begin{equation*} 
    \mR \triangleq \mT^{-1} + \mS^{\herm} \mS 
    \qquad\text{and}\qquad
    \vz \triangleq \mS^{\herm} \vy_{\mathcal{F}}
\end{equation*}
}%
for clarity. 
\cref{eq:approximate_collapsed_likelihood} requires computing the determinant \(\det\left(\mR\right)\) and quadratic form \(\vz^{\herm} \mR^{-1} \vz\), which is not trivial even with stripe matrices.
For this, \citet{ladaycia_performance_2017} proposed to use the Schur complement recursively, which, for \(\mR \in \mathbb{C}^{k N^{\prime} \times k N^{\prime}}\) comprised of \(N^{\prime} \times N^{\prime}\) blocks, yields its inverse \(\mR^{-1}\) and determinant \(\det \mR\) in \(\mathcal{O}\left(k^3 N\right)\) time.
Unfortunately, the recursive Schur complement is numerically unstable.
Instead, since \(\mR\) is always PD as \(\mT\) is PD, we can consider decompositions specialized for PD matrices such as the Cholesky or LDL\(^{\top}\) decompositions.
In our case, we observed that LDL\(^{\top}\) was numerically more stable.
For a PD matrix such as \(\mR\), the LDL\(^{\top}\) decomposition yields the unique lower triangular matrix \(\mL \in \mathbb{C}^{k N^{\prime} \times k N^{\prime}}\) and diagonal matrix \(\mD \in \mathbb{C}^{k N^{\prime} \times k N^{\prime}}\) such that
{%
\setlength{\belowdisplayskip}{.5ex} \setlength{\belowdisplayshortskip}{.5ex}
\setlength{\abovedisplayskip}{.5ex} \setlength{\abovedisplayshortskip}{.5ex}
\begin{equation*}
    \mR = \mL \mD \mL^{\herm}.
\end{equation*}
}%
Conveniently, for stripe matrices, computing the block-LDL\(^{\top}\) decomposition~\citep[Algorithm 4.1]{fang_stability_2011}, shown in~\cref{alg:ldl}, yields the full LDL\(^{\top}\) decomposition (\cref{thm:blockldl_is_ldl} in \cref{section:appendix_stripe}), where \(\mL\) has the same stripe matrix structure as \(\mR\) in the lower triangular region.
This takes \(\mathcal{O}(k^3 N)\) operations.

\begin{figure}[t]
  \removelatexerror
  {
  \begin{algorithm2e}[H]
    \DontPrintSemicolon
    \SetAlgoLined
    \KwIn{Stripe matrix \(\mR \in \mathbb{C}^{k N^{\prime} \times k N^{\prime}}\) with \(N^{\prime} \times N^{\prime}\) blocks.
    }
    \For{ \( i = 1, \ldots, k\) }{
      \(\mD_{ii} = \mR_{ii} - {\textstyle\sum^{i-1}_{k=1}} \mL_{ik} \mD_{kk}\mL_{ik}^{*}\)\;
      \For{ \( j = 1, \ldots, k\) }{
{%
\setlength{\belowdisplayskip}{.5ex} \setlength{\belowdisplayshortskip}{.5ex}
\setlength{\abovedisplayskip}{.5ex} \setlength{\abovedisplayshortskip}{.5ex}
\begin{alignat*}{2}
    \hspace{-1em}
    \mL_{ij} &=  \begin{cases}
        \boldupright{I}_{N^{\prime}} &\text{if \(i = j\)}
        \\
        \left( \mR_{ij} - {\textstyle\sum^{j-1}_{k=1}} \mL_{ik} \mD_{kk} \mL_{jk}^{*} \right) \, \mD_{ii}^{-1} 
        &\text{if \(i > j\)} 
        \\
        \boldupright{O}_{N^{\prime}}
        &\text{else}
    \end{cases}
\end{alignat*}
}%
  \vspace{-1ex}
        }
    }
    \caption{Block-LDL\(^{\top}\) Decomposition}\label{alg:ldl}
  \end{algorithm2e}
  }
  \vspace{-3ex}
\end{figure}

Equipped with the LDL\(^{\top}\) decomposition of \(\mR\), we can now compute the quadratic form and determinants as
{%
\setlength{\belowdisplayskip}{1ex} \setlength{\belowdisplayshortskip}{1ex}
\setlength{\abovedisplayskip}{1ex} \setlength{\abovedisplayshortskip}{1ex}
\begin{equation}
    \vz^{\herm} \mR^{-1} \vz 
    =
    \norm{\mL^{-1} \vz}_{\mD^{-1}}^2 , \;
    \mathrm{det}\left(\mR\right)
    =
    {\textstyle \prod^{k}_{j=1} }\det\left(\mD_{jj}\right).
\end{equation}
}%
Similarly to~\cref{alg:ldl}, \(\mL^{-1} \vz\) can be computed through a block forward-substitution and the determinant of a diagonal matrix such as \(\mD_{jj}\) is the product of the diagonal.
This now completes computing \cref{eq:approximate_collapsed_likelihood}.
Whenever the LDL\(^{\top}\) decomposition fails, or the resulting likelihood is not strictly positive due to numerical errors, we set the likelihood as \(0\) such that RJMCMC can reject the offending proposal.

\begin{figure}[t]
  \removelatexerror
  {
  \begin{algorithm2e}[H]
    \DontPrintSemicolon
    \SetAlgoLined
    \KwIn{previous state \((k, \vtheta_k, v)\) and \newline
    update move probability \(\tau \in (0, 1)\).
    }
    \(u \sim \mathsf{Uniform}\left[0, 1\right]\)\;
    \eIf{\(u < \tau\)}{
        Execute update move
    }{
    \(k^{\prime} = k + v\)\;
    \(\vu_{k \mapsto k^{\prime}} \sim q_{k \mapsto k^{\prime}}\left(\vu_{k \mapsto k^{\prime}}\right)\) \;
    \( \left( \vtheta_{k^{\prime}}, \vu_{k^{\prime} \mapsto k} \right) \leftarrow \mathcal{T}_{k \mapsto k^{\prime}}\left( \vtheta_k, \vu_{k \mapsto k^{\prime}} \right) \) \;
    Compute \(\alpha \leftarrow \min(1, r)\) with \(r\) in \cref{eq:nrjmcmc_accept_reject}\;
    Accept or reject \((k^{\prime}, \vtheta_{k^{\prime}})\) with probability \(\alpha\)\;
    \If{ rejected }{
        \(v \leftarrow -v\)\;
    }
    }
    \caption{Non-Reversible Jump MCMC Transition}\label{alg:nrjmcmc}
  \end{algorithm2e}
  }
  \vspace{-4ex}
\end{figure}

\paragraph{Computational Complexity}
The main bottleneck of computing \cref{eq:approximate_collapsed_likelihood} is computing the LDL\(^{\top}\) decomposition, which takes \(\mathcal{O}(N k^3)\) operations.
The forward substitution also takes \(\mathcal{O}(N k^2)\) operations.
Since \(\vy_{\mathcal{F}}\) only needs to be computed once, the cost of the DFT is fully amortized.
The products \(\mS^{\herm} \mS\) and \(\mS^{\herm} \vy_{\mathcal{F}}\) take \(\mathcal{O}(k^3 N)\) and \(\mathcal{O}(N M k)\) operations, respectively.
Therefore, the overall cost is \(\mathcal{O}( N k^3 + N M k)\).
Note that, \(k \leq k_{\text{max}} < M\).
Therefore, the worst-case complexity can be summarized to \(\mathcal{O}( N M^3)\).
This is comparable to the cost of computing the likelihood of time--frequency models \citep[Eq. 14.99, Eq. 14.102]{chung_doa_2014}, which is \(\mathcal{O}(B M^2 k)\), where the pseudoinverse of the \(M \times k\) ``steering matrix'' is computed for each of the \(B\) frequency bins.

\vspace{-1.5ex}
\subsection{Inference via Non-Reversible Jump MCMC}\label{section:rjmcmc}
Now, we will discuss how to draw samples from the posterior in \cref{eq:posterior}.
Let us denote a single collection of parameters for a sample as \(\vtheta_{k} = \left(\vphi_{1\text{:}k}, \vgamma_{1\text{:}k}\right) \in \mathbb{R}^{d_k}\) such that the target posterior is also denoted \(\pi\left(k, \vtheta_k\right) : \cup_{k \in \mathcal{K}} \{k\} \times \mathbb{R}^{d_k} \to \mathbb{R}_{\geq 0}\), where \(\mathcal{K} = [k_{\mathrm{max}}]\).
Unlike typical Bayesian analysis setups, the dimension of our posterior changes depending on \(k\).
\textit{Reversible jump  Markov chain Monte Carlo} (RJMCMC;~\citep{green_reversible_1995,green_transdimensional_2003,hastie_model_2012,sisson_transdimensional_2005}) is a family of MCMC algorithms~\citep{robert_monte_2004} specialized for such setting.
In particular, we use the birth-death-update process~\citep{green_reversible_1995} with its non-reversible extension proposed by~\citet{gagnon_nonreversible_2021}.

\begin{figure*}[t]
    \vspace{-5ex}
    \centering
    \subfloat[SNR = -4 dB]{
        \includegraphics[scale=1.0]{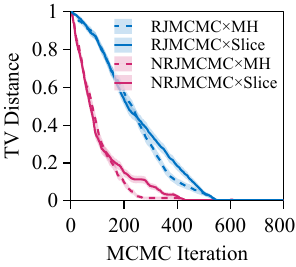}
    }
    \subfloat[SNR = 0 dB]{
        \includegraphics[scale=1.0]{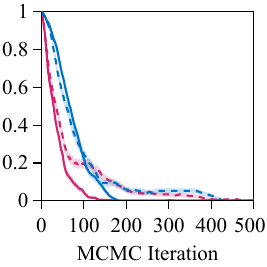}
    }
    \subfloat[SNR = 4 dB]{
        \includegraphics[scale=1.0]{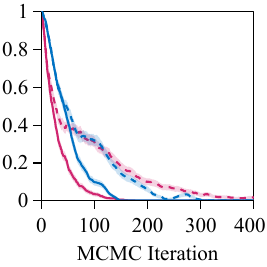}
    }
    \caption{
    \textbf{Convergence of the Model Order ($k$) Chain in TV Distance Under Varying SNR Levels.}
    The solid lines are the estimated TV distances, while the colored bands are the corresponding \(95\%\) bootstrap confidence intervals.
    }\label{fig:rjmcmc_eval}
    \vspace{-2ex}
\end{figure*}

\paragraph{Reversible Jump MCMC}
Traditional RJMCMC~\citep{green_reversible_1995} operates analogously to Metropolis-Hastings~\citep{metropolis_equation_1953,hastings_monte_1970}, where, given a state of the Markov chain \((k, \vtheta_k)\), a transition to a different model space \(k^{\prime}\) may be proposed as
{%
\setlength{\belowdisplayskip}{.5ex} \setlength{\belowdisplayshortskip}{.5ex}
\setlength{\abovedisplayskip}{.5ex} \setlength{\abovedisplayshortskip}{.5ex}
\begin{equation*}
    k^{\prime} \sim q\left(k, k^{\prime}\right).
\end{equation*}
}%
When \(k \neq k^{\prime}\), the proposal ``jumps'' to a different model space \(k^{\prime}\).
Therefore, such a proposal is referred to as a jump proposal and \(q\left(k, k^{\prime}\right)\) is referred to a jump proposal kernel.
The parameters in the model space of \(k^{\prime}\) of the jump proposal are \(\vtheta^{\prime}\) computed as
{%
\setlength{\belowdisplayskip}{.5ex} \setlength{\belowdisplayshortskip}{.5ex}
\setlength{\abovedisplayskip}{.5ex} \setlength{\abovedisplayshortskip}{.5ex}
\begin{equation*}
    \left( \vtheta_{k^{\prime}}, \vu_{k^{\prime} \mapsto k} \right) = \mathcal{T}_{k \mapsto k^{\prime}}\left( \vtheta_k, \vu_{k \mapsto k^{\prime}} \right),
\end{equation*}
}%
where \(\vu_{k \mapsto k^{\prime}}\) are the auxiliary variables needed to enable the jump, which are sampled as
{%
\setlength{\belowdisplayskip}{1ex} \setlength{\belowdisplayshortskip}{1ex}
\setlength{\abovedisplayskip}{1ex} \setlength{\abovedisplayshortskip}{1ex}
\begin{equation*}
    \vu_{k \mapsto k^{\prime}} \sim q_{k \mapsto k^{\prime}}\left(\vu_{k \mapsto k^{\prime}}\right).
\end{equation*}
}%
The proposal \((k^{\prime}, \vtheta_{k^{\prime}})\) is then accepted or rejected according to a Metropolis-Hastings step (MH) with probability
{ %
\setlength{\belowdisplayskip}{1ex} \setlength{\belowdisplayshortskip}{1ex}
\setlength{\abovedisplayskip}{1ex} \setlength{\abovedisplayshortskip}{1ex}
\begin{equation*}
    \alpha\left(\left(k^{\prime}, \vtheta_{k^{\prime}}\right), \left(k, \vtheta_k\right)\right) = \min\left(1, r\right)
\end{equation*}
}%
according to the Metropolis-Hastings-Green (MHG) ratio 
{ %
\setlength{\belowdisplayskip}{1ex} \setlength{\belowdisplayshortskip}{1ex}
\setlength{\abovedisplayskip}{1ex} \setlength{\abovedisplayshortskip}{1ex}
\begin{align*}
  r
  =
  \frac{ 
    \pi\left(k^{\prime}, \vtheta_{k^{\prime}}\right)
  }{ 
    \pi\left(k, \vtheta_{k}\right) 
  }
  \frac{ q\left(k^{\prime}, k\right) }{ q\left(k, k^{\prime}\right) }
  \frac{ 
    q_{k^{\prime} \mapsto k}\left(\vu_{k^{\prime} \mapsto k}\right)
  }{ 
    q_{k \mapsto k^{\prime}}\left(\vu_{k \mapsto k^{\prime}}\right)
  }
  {J_{k \mapsto k^{\prime}}\left(\vtheta_{k}, \vu_{k \mapsto  k^{\prime}}\right)},
\end{align*}
}%
where \(J_{k \mapsto k^{\prime}}\) denotes the absolute determinant of the Jacobian of \(\mathcal{T}_{k \mapsto k^{\prime}}\).
Since we use the birth-death-update process, which only adds or removes elements, \(J_{k \mapsto k^{\prime}} = 1\).

\paragraph{Non-Reversible Jump MCMC}
Over the years, it has been shown that simulating non-reversible processes often results in faster mixing and lower autocorrelation~\citep{diaconis_analysis_2000}.
For the RJMCMC setting, \citet{gagnon_nonreversible_2021} proposed to apply a technique known as \textit{lifting}, which augments the Markov chain with a binary random variable \(v \in \{+1, -1\}\) such that the Markov chain is now \( {\{ (k^t, \vtheta_{k^t}^t, v^t) \} }_{t \geq 0}\).
In the context of RJMCMC, at each MCMC step, the jump proposal is deterministically chosen as \(k^{\prime} = k + v\), and we proceed with the MH step with the ratio
{ %
\setlength{\belowdisplayskip}{1ex} \setlength{\belowdisplayshortskip}{1ex}
\setlength{\abovedisplayskip}{1ex} \setlength{\abovedisplayshortskip}{1ex}
\begin{align}
  r
  =
  \frac{ 
    \pi\left(k^{\prime}, \vtheta_{k^{\prime}}\right)
  }{ 
    \pi\left(k, \vtheta_{k}\right) 
  }
  \frac{ 
    q_{k^{\prime} \mapsto k}\left(\vu_{k^{\prime} \mapsto k}\right)
  }{ 
    q_{k \mapsto k^{\prime}}\left(\vu_{k \mapsto k^{\prime}}\right)
  }
  J_{k \mapsto k^{\prime}}\left(\vtheta_k, \vu_{k \mapsto  k^{\prime}}\right)
  .
  \label{eq:nrjmcmc_accept_reject}
\end{align}
The ``jump direction'' \(v\) is flipped whenever the proposal is rejected.
The resulting chain is a piecewise-deterministic Markov process (PDMP), which is non-reversible, hence non-reversible jump MCMC (NRJMCMC; \cref{alg:nrjmcmc}), but still preserves the correct stationary distribution.
The main benefit of NRJMCMC is lower autocorrelation than RJMCMC.
Under idealized conditions,~\citet{gagnon_nonreversible_2021} show that NRJMCMC can achieve significantly lower autocorrelation than RJMCMC when the model order posterior is not too concentrated.
On the other hand, in the worst case, NRJMCMC achieves an autocorrelation that is no worse than twice that of RJMCMC \citep{gagnon_theoretical_2026}.
Therefore, even when NRJMCMC fails, it should perform comparably to RJMCMC.

\paragraph{Birth-Death-Update Process}
For the jump moves, we use the classic birth-death-update process~\citep{green_reversible_1995}.
It alternates between the \textit{birth}, \textit{death}, and \textit{update} moves, which correspond to the jump move proposed when \(k^{\prime} = k+1\), \(k^{\prime} = k-1\), and \(k^{\prime} = k\) respectively.
The birth move increases the model order through \textit{insertion}:
{%
\setlength{\belowdisplayskip}{1ex} \setlength{\belowdisplayshortskip}{1ex}
\setlength{\abovedisplayskip}{0ex} \setlength{\abovedisplayshortskip}{0ex}
\begin{align*}
    \vphi_{1\text{:}k} \oplus_j \phi^{\prime}
    =
    \left( \phi_1, \ldots, \phi_{j-1}, \phi^{\prime}, \phi_{j}, \ldots, \phi_k\right),
\end{align*}
}%
while the death move decreases it through \textit{removal} 
{%
\setlength{\belowdisplayskip}{.5ex} \setlength{\belowdisplayshortskip}{.5ex}
\setlength{\abovedisplayskip}{0ex} \setlength{\abovedisplayshortskip}{0ex}
\begin{align*}
    \vphi_{1\text{:}k,-j} 
    =
    \left( \phi_1, \ldots, \phi_{j-1}, \phi_{j+1}, \ldots, \phi_k\right),
\end{align*}
}%
and the update move updates the parameters through a regular MCMC kernel without altering the model order.

\paragraph{Birth Move}
During the birth move, a \textit{newborn} \((\phi^{\prime}, \gamma^{\prime})\) and its insertion position \(j\) are proposed as
{%
\setlength{\belowdisplayskip}{0.5ex} \setlength{\belowdisplayshortskip}{0.5ex}
\setlength{\abovedisplayskip}{0.5ex} \setlength{\abovedisplayshortskip}{0.5ex}
\begin{align*}
  \phi^{\prime} \sim q\left(\phi\right),\quad
  \gamma^{\prime} \sim q\left(\gamma\right),\quad
  j \sim \mathsf{Uniform}\left\{0, \ldots, k\right\},
\end{align*}
}%
forming the auxiliary variable \(\vu_{k \mapsto k+1} = \left(\phi^{\prime}, \gamma^{\prime}, j\right)\).

The birth proposal is then
\begin{align*}
  \vtheta_{k+1}
  = \left(
  \vphi_{1\text{:}k} \oplus_j \phi^{\prime},\;\;
  \vgamma_{1\text{:}k} \oplus_j \gamma^{\prime}
 \right)
  = \mathcal{T}_{k \mapsto k+1}\left(\vtheta_k, \vu_{k \mapsto k+1}\right).
\end{align*}
The probability densities of the auxiliary variables for the forward and backward jumps of the birth move are given as
{%
\setlength{\belowdisplayskip}{0.5ex} \setlength{\belowdisplayshortskip}{0.5ex}
\setlength{\abovedisplayskip}{0.5ex} \setlength{\abovedisplayshortskip}{0.5ex}
\begin{alignat*}{2}
  q_{k \mapsto k+1}\left(\vu_{k \mapsto k+1}\right) &= \frac{1}{k + 1} q\left(\phi^{\prime}, \gamma^{\prime}\right), &&\quad\text{(Birth Forward)} \\
  q_{k+1 \mapsto k}\left(\vu_{k+1 \mapsto k}\right) &= \frac{1}{k + 1}. &&\quad\text{(Birth Reverse)}
\end{alignat*}
}%
Previously, a common practice following \citep{andrieu_joint_1999} was to insert the newborn at the end of the vector.
As demonstrated by~\citet{roodaki_comments_2013}, this breaks the symmetry between the birth and death moves, resulting in a biased RJMCMC procedure, hence the involvement of \(j\) here.

\paragraph{Death Move}
For the death move, we only have to select the index of the removal candidate 
{%
\setlength{\belowdisplayskip}{0.5ex} \setlength{\belowdisplayshortskip}{0.5ex}
\setlength{\abovedisplayskip}{0.5ex} \setlength{\abovedisplayshortskip}{0.5ex}
\begin{align*}
  j \sim \mathsf{Uniform}\left\{1, \ldots, k\right\},
\end{align*}
}%
where it is then removed as
{%
\setlength{\belowdisplayskip}{0.5ex} \setlength{\belowdisplayshortskip}{0.5ex}
\setlength{\abovedisplayskip}{0.5ex} \setlength{\abovedisplayshortskip}{0.5ex}
\begin{align*}
   \vtheta_{k-1}
   = 
   \left(
     \vphi_{1\text{:}k,-j},\;\;
     \vgamma_{1\text{:}k,-j}
   \right)
   =
   \mathcal{T}_{k \mapsto k-1}\left( \vtheta_{k}, \vu_{k \mapsto k-1} \right).
\end{align*}
}%
The auxiliary variable is then \(\vu_{k \mapsto k-1} = j\).
The densities for the forward and backward auxiliary variables are given as
\begin{alignat*}{2}
  q_{k \mapsto k-1}\left(\vu_{k \mapsto k-1}\right) &= \frac{1}{k}, &&\quad\text{(Death Forward)} \\
  q_{k-1 \mapsto k}\left(\vu_{k-1 \mapsto k}\right) &= \frac{1}{k} q\left(\phi_j, \gamma_j\right). &&\quad\text{(Death Reverse)}.
\end{alignat*}

\paragraph{Update Move}
For the update move, we apply slice sampling~\citep{neal_slice_2003}.
Compared to independent or random-walk MH~\citep{hastings_monte_1970} used in previous works~\citep{ng_wideband_2005,larocque_reversible_2002,copsey_bayesian_2002,davy_bayesian_2006,eches_estimating_2010,davy_classification_2002,liu_bayesian_2022,amrouche_efficient_2022}, it achieves remarkable performance both in practice and in theory. (See \citep{power_weak_2024} and references therein.)
Furthermore, the posterior of the DoAs is multi-modal, making gradient-based samplers ineffective.
In contrast, slice sampling can handle multi-modality to some degree as long as the ``search interval'' $w$ is chosen to be wide enough~\citep[\S 4.2]{power_weak_2024}.
In this work, we use the ``stepping-out'' variant \citep[Scheme 3]{neal_slice_2003} with \(w = 2\), which is augmented into a coordinate-wise Gibbs scheme~\citep{geman_stochastic_1984,gelfand_samplingbased_1990} with random permutation scans~\citep[Alg. 42]{robert_monte_2004}.
The update move selection probability is set as \(\tau = 0.1\).
We empirically compare the performance of slice sampling against Metropolis-Hastings samplers in \cref{section:mixing}.

\paragraph{Convergence}
Let us now comment on the ergodicity/mixing/convergence properties of the considered MCMC procedures.
If, for any $k \geq 0$, the SNRs $\gamma_{1:k}$ are fixed, then our model becomes structurally identical to that in \citep{andrieu_joint_1999}, where the RJMCMC chain is shown to be uniformly geometrically ergodic under the mild condition that $\vy$ is noisy enough \citep[Thm. 1]{andrieu_joint_1999}.
Although fixing $\gamma_{1:k}$ is less realistic, this does suggest that the multi-modality of the posterior of $\phi_{1\text{:}k}$ is not a problem; a mode can be escaped via the death move.
For general setups, RJMCMC is geometrically ergodic as long as the update moves are also geometrically ergodic~\citep{qin_geometric_2024}, where stepping-out slice sampling has been shown to be geometrically ergodic even on certain multi-modal targets as long as $w$ is large enough~\citep[\S 4.2]{power_weak_2024}.
Meanwhile, the ergodicity of NRJMCMC is not very well understood due to the fundamental difficulty of studying non-reversible Markov chains.
Therefore, we empirically evaluate convergence in \cref{section:mixing}.
Obtaining more general theoretical guarantees is an active research topic beyond the scope of this paper.

\begin{figure*}[t]
    \vspace{-8ex}
    \centering
    \includegraphics[scale=1.0]{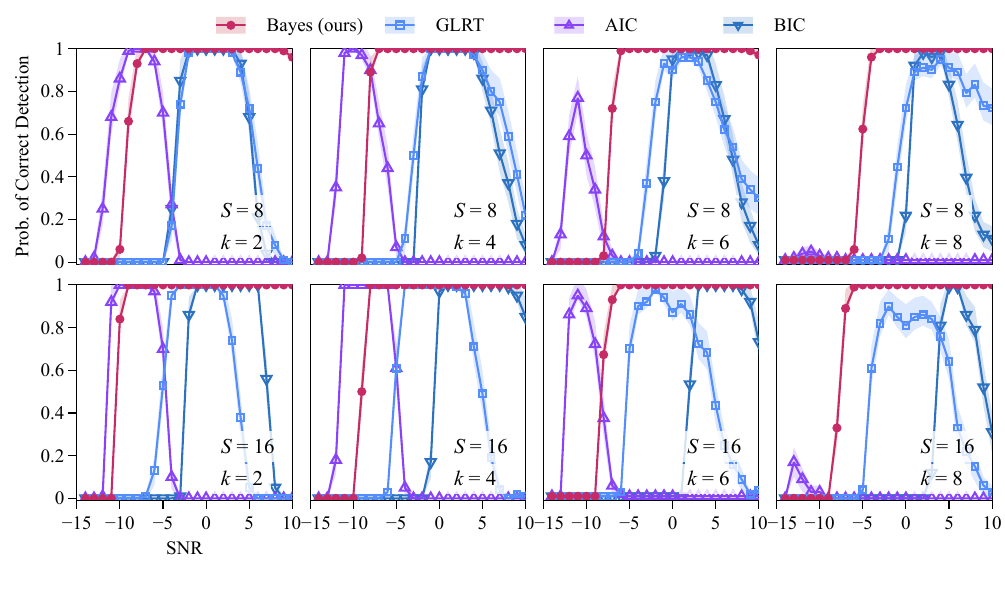}
    \vspace{-4ex}
    \caption{\textbf{
    Wideband Signal Detection Performance Versus SNR, \# of Snapshots (\(S\)), and \# of Targets (\(k\)).}
    The solid lines are the mean estimated from 100 replications, while the colored bands are \(95\%\) bootstrap confidence intervals of the mean.
    }\label{fig:detection_wideband}
    \vspace{-3ex}
\end{figure*}

\vspace{-2ex}
\section{Evaluation}\label{section:evaluation}
\vspace{-1ex}
\subsection{Experimental Setup}
\paragraph{Implementation}
We implemented our proposed scheme in the Julia language~\citep{bezanson_julia_2017}.
The probability distributions were provided by the \texttt{Distributions.jl} library \citep{lin_juliastats_2024} and the 
stripe matrix operations were implemented using the \texttt{Tullio.jl}~\citep{abbott_mcabbott_2023} array programming framework.
The correctness of our RJMCMC implementation was verified through the tests proposed in \citep{gandy_unit_2021}.
For the auxiliary jump variable proposal, we set
\begin{align*}
    q\left(\phi\right) &= \mathsf{Uniform}\left(\phi; \left[-\nicefrac{\pi}{2}, \nicefrac{\pi}{2}\right]\right) \\
    q\left(\gamma\right) &= \mathsf{LogNormal}\left(\gamma; 0, 2\right) \; .
\end{align*}
To ensure \(\gamma_j\) stays within \(\mathbb{R}_{>0}\), we sample it in log space and apply a Jacobian adjustment to the joint likelihood.
For the remaining hyperparameters, we set \(\alpha = \beta = 0\).
Then $p(\sigma^2)$ corresponds to Jeffrey's scale invariant prior~\citep{consonni_prior_2018}.
This means that our overall system is invariant to signal power.
The maximum number of targets is \(k_{\max} = M - 1\).
For the detection experiments, we draw \(2^{12}\) posterior samples with NRJMCMC after discarding \(2^{10}\) samples as burn-in and apply the procedure in \cref{section:detection_estimation_procedure}.
All of the code used in this work is publicly available\footnote{\textsc{GitHub} link: \url{https://github.com/Red-Portal/WidebandDoA.jl}}.

\paragraph{Baselines}
We consider the following baselines: 
\begin{itemize}
    \item \textbf{GLRT}:
    This is a GLRT-based detection method proposed by~\citet{chung_detection_2007}.
    The \(p\)-values are obtained by bootstrapping~\citep{zoubir_bootstrap_2001} over the frequency bins, where multiple testing is performed according to the Benjamini-Hochberg procedure~\citep{benjamini_controlling_1995}, which controls the false discovery rate \(q \in (0, 1)\).
    Following~\citep{chung_detection_2007}, we choose \(q = 0.1\).
    
    \item \textbf{AIC, BIC}: These are the classical information theoretic approaches as proposed by \citet{akaike_new_1974} using the AIC and BIC criteria, respectively~\citep{stoica_modelorder_2004,ding_model_2018}.
    The BIC criterion was also commonly referred to as the minimum description length (MDL) criterion in the signal processing literature.
\end{itemize}
Both approaches require solving a series of maximum likelihood problems over the range of models \(k = 1\) up to \(k = k_{\mathrm{max}}\)
We give a slight advantage to the baselines by narrowing down the range to \(k_{\mathrm{max}} = 10\) instead of \(M - 1\) as for RJMCMC.
We use the likelihood of the deterministic signal model~\citep[\S 3.14.5.1.1, Eq. 14.99]{chung_doa_2014}, which is maximized through the space alternating generalized expectation maximization (SAGE) algorithm~\citep{cadalli_wideband_1999} analyzed in~\citep{chung_comparative_2001}.
(The general SAGE algorithm was originally proposed in~\citep{fessler_spacealternating_1994}.)
To deal with multimodality, we initialize each linesearch with DIRECT~\citep{jones_lipschitzian_1993} followed by a refinement with L-BFGS~\citep{liu_limited_1989}.

\begin{figure*}[t]
    \vspace{-8ex}
    \centering
        \includegraphics[scale=1.0]{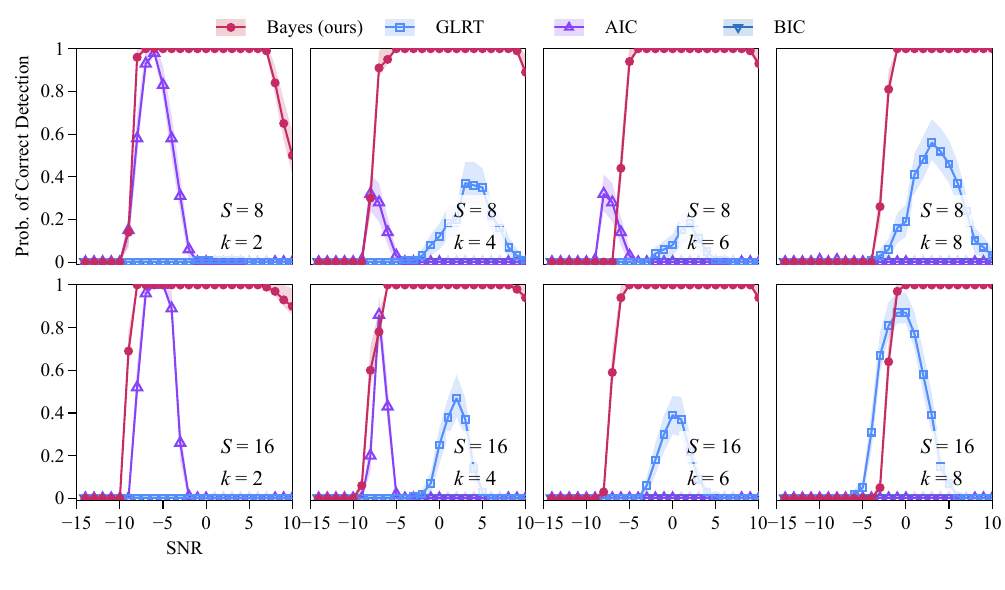}
    \vspace{-4ex}
    \caption{\textbf{Narrowband signal detection performance versus SNR, \# of snapshots (\(S\)), and \# of targets (\(k\)).}
    The solid lines are the mean estimated from 100 replications, while the colored bands are \(95\%\) bootstrap confidence intervals of the mean.
    }\label{fig:detection_narrowband}
    \vspace{-3ex}
\end{figure*}

\paragraph{Simulation Methodology}
For each \(j \in [k]\) for a true \(k\), we generate \(\vx_j\) by band-limiting white Gaussian noise to have a signal power of \(P_j = \mathbb{E}\vx_j^2\).
The signal is then delayed using the FIR delay filter by \citet{pei_closed_2014}.
Lastly, white Gaussian noise \(\veta\) with power \(\sigma^2\) is added such that the \(j\)th source signal has \(\mathrm{SNR} = 10 \log_{10} P_j/\sigma^2\) [dB].
For ML-based methods that use the time--frequency model, the received signal \(\vy\) is STFT-ed with \(B = 32\) frequency bins with a rectangular window and no overlap.
Since we only use real-valued signals, a one-sided STFT is applied, resulting in 16 frequency bins after excluding the DC bin.
The sampling frequency is \(f_{s} = 3\) kHz, the array is a ULA with 20 sensors, the sensor spacing is 0.5 m, and the propagation speed is \(c = 1500\) m/s.

\vspace{-2ex}
\subsection{Mixing of the RJMCMC Chains}\label{section:mixing}
Since our procedure crucially depends on RJMCMC, we first empirically verify the mixing of our RJMCMC chains and evaluate the design choices made in \cref{section:rjmcmc}.
Our evaluation has two main axes: Reversible jump moves (RJMCMC) versus non-reversible jump moves (NRJMCMC), and for the update move, slice sampling (Slice) versus Metropolis-Hastings samplers (MH).

\paragraph{Setup}
To evaluate the mixing of the chains, we will monitor the total variation (TV) distance between the marginal RJMCMC kernel and the true posterior of the model order ($k$) chain.
We estimate the true model order posterior by simulating a long RJMCMC chain of $2^{18}$ samples, where the first $2^{17}$ samples were discarded as burn-in.
The marginal probability of the RJMCMC kernel is estimated by simulating $2^{10}$ independent and identically distributed Markov chains.
Since the model order chain operates on a discrete state space, the TV distance can be computed in a closed form.

For the MH baselines, following \citep{andrieu_joint_1999}, the proposals were set as mixtures between independent and random walk proposals.
The independent proposal ensures that the chain doesn't get stuck in a local mode, while the random walk proposal ensures that the chain mixes well locally.
For $\phi_j$, we use a mixture of a uniform proposal over $\left[-\nicefrac{\pi}{2}, \nicefrac{\pi}{2}\right]$ and a Gaussian random walk proposal with variance $0.1^2$.
Meanwhile, for $\log \gamma_j$ (operating in log space), we use an independent proposal $\mathsf{Normal}(0, 2^2)$ and a Gaussian random walk proposal with variance $0.5^2$.
In all cases, the mixture weight is set to $0.2$ for the independent and $0.8$ for the random walk proposal.

For the data, we simulated two targets at $\phi_{1} = -\nicefrac{\pi}{2}$ and $\phi_2 = \nicefrac{\pi}{2}$, respectively, each emitting a signal equally-powered over $[10 \text{Hz}, 1 \text{kHz}]$ with length $N = 128$ and varying SNRs.

\paragraph{Results}
The results are shown in \cref{fig:rjmcmc_eval}.
In all cases, the chain converged within 600 iterations.
When comparing reversible (RJMCMC) versus non-reversible jump proposals (NRJMCMC), we see that non-reversibility prevails in -4 dB, whereas in 0 and 4 dB, the choice of update move kernel (Slice versus MH) is more important.
In particular, at 4 dB, using Slice instead of MH is decisive for both RJMCMC and NRJMCMC.
Overall, we conclude that NRJMCMC with slice sampling leads to good performance in scenarios.
In particular, it is well known that random walk MH is sensitive to tuning, and a choice of random walk variance optimal in one scenario will not be adequate in different conditions.
On the other hand, slice sampling is relatively insensitive to tuning, leading to robust performance even across varying SNRs.

\vspace{-1.5ex}
\subsection{Wideband Signal Detection Performance}\label{section:detection_wideband}

From now on, we will evaluate the detection performance of our procedure.

\paragraph{Setup}
We first consider equally spaced, equally powered wideband signals with varying snapshots, targets, and SNR.
For our Bayesian procedure (Bayes), which does not involve the STFT, we use the equivalent number of \(N = B \cdot S\) samples.
The bandwidths of the source signals were set to be [10 Hz, 1 kHz], which is slightly less than the half-wavelength constraint, making the problem more challenging.

\paragraph{Results}
The results are shown in \cref{fig:detection_wideband}.
AIC generally performs poorly due to the well-known issue of overestimating the number of sources.
In contrast, GLRT and BIC maintain high accuracy for a wider range of SNRs, where GLRT is better in low SNRs, while BIC is better at high SNRs.
Compared to these, our Bayesian detection procedure exhibits good low-SNR performance while maintaining high accuracy for the widest range of SNRs.

\vspace{-1.5ex}
\subsection{Narrowband Signal Detection Performance}\label{section:detection_narrowband}
\vspace{-1ex}
\paragraph{Setup}
Next, we evaluate the performance of wideband detection methods faced with narrowband signals.
In this setting, our prior on the source signals is grossly misspecified since, for each $k$, the Gaussian prior assumes $\vx_{j}$ has a flat spectrum conditional on $\gamma_j$.
Similarly to \cref{section:detection_wideband}, we consider equally powered equally spaced signals, where the bandwidth is set as [500 Hz, 600 Hz], which is close to a single frequency bin.

\begin{figure}[t]
    \vspace{-2ex}
    \centering
    \includegraphics[scale=1]{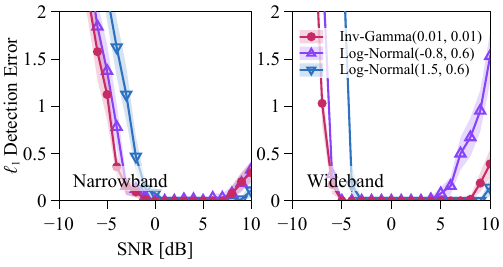}
    \caption{
    \textbf{Comparison of SNR Priors.}
    For a description of the considered priors, refer to \cref{section:snr_priors}.
    The solid lines are the mean estimated from $128$ replications, while the colored bands are \(95\%\) bootstrap confidence intervals of the mean.
}\label{fig:snr_priors}
    \vspace{-3ex}
\end{figure}


\paragraph{Results}
The results are shown in \cref{fig:detection_narrowband}.
For narrowband signals, information-theoretic methods perform much poorly.
This is because the complexity penalty penalizes the complexity of modeling all the frequency bins, while the improvement in likelihood comes from only a single bin.
In contrast, Bayes showed good performance for a wide range of SNR.
However, unlike the wideband case, it required a higher SNR to transition to correctly detecting the targets.
Meanwhile, GLRT underperformed compared to Bayes.
While the performance of GLRT shown here is worse compared to the performance reported in \citep{chung_detection_2007}, this is because, unlike in \citep{chung_detection_2007}, we applied the wideband version of GLRT to a narrowband signal: the bins not containing the target signal reduce the overall power of the test procedure.

\vspace{-1.5ex}
\subsection{Comparison of SNR Priors}\label{section:snr_priors}
\vspace{-1ex}
\paragraph{Setup}
We evaluate the effect of the choice of SNR prior $p(\gamma_j)$ on the detection performance on both wideband (full bandwidth) and narrowband (100 Hz bandwidth) signals.
We compare our choice of uninformative prior $\text{Inv-Gamma}(0.01, 0.01)$, which mimics Jeffrey's scale invariant prior, against informative priors, $\text{Log-Normal}(-0.8, 0.6)$ and $\text{Log-Normal}(1.5, 0.6)$, which respectively put most mass around -5 dB and 5 dB. (The log-normal priors correspond to putting a Gaussian prior in dB scale.)
We simulate $N = 128$ measurements emitted from $k = 5$ targets.

\paragraph{Results}
The results are shown in \cref{fig:snr_priors}.
We can see that $\text{Log-Normal}(-0.8, 0.6)$ achieves lower error in lower SNRs and higher error in higher SNRs, $\text{Log-Normal}(1.5, 0.6)$ exhibits opposite behavior.
In contrast, the uninformative prior $\text{Inv-Gamma}(0.01, 0.01)$ is close to the best of both worlds; its low and high SNR performance is close to the best performing prior in each regime.
Therefore, $\text{Inv-Gamma}(0.01, 0.01)$ should serve as a good default option.

\vspace{-1.5ex}
\subsection{Computation Cost}\label{section:computational_cost}
\paragraph{Setup}
We now empirically evaluate the computational cost of our method compared to the baselines.
To ensure a fair comparison, we set $k_{\mathrm{max}} = M - 1$ for the baselines, which is how it would be set in practice.
All methods use $S = 8$ snapshots, while the experiments were conducted on a workstation with an Intel i9-11900F CPU.

\begin{figure}[t]
    \vspace{-2ex}
    \centering
    \includegraphics[scale=1.0]{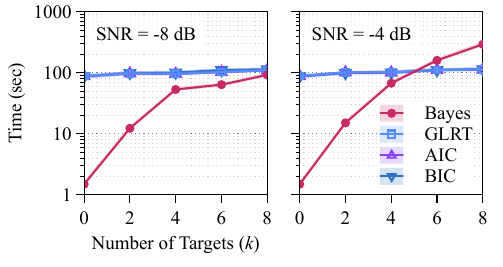}
    \caption{
    \textbf{Comparison of Execution Time.}
    The solid lines are the mean estimated from $100$ replications, while the colored bands are \(95\%\) bootstrap confidence intervals of the mean.
}\label{fig:execution_time}
    \vspace{-1ex}
\end{figure}

\paragraph{Results}
Recall the time complexity of our detection scheme discussed in paragraph d) of \cref{section:likelihood_computation}.
The computational cost of our method tends to scale poorly in terms of $k$ due to the $\mathcal{O}(k^3)$ complexity.
This fact can be noticed in \cref{fig:execution_time}, where our method is most efficient for smaller values of $k$.
Also, our method is more efficient on lower SNRs ($-8$ dB) than higher SNRs ($-4$ dB).
This is because higher uncertainty at low SNRs manifests as wider posterior modes, resulting in fewer rejections.

\section{Discussions}\label{section:conclusion}
In this work, we have developed a new model for wideband signals that enables efficient, fully Bayesian signal detection.
Experimentally, our Bayesian approach achieves higher detection probability for a wider range of SNRs compared to frequentist and information-theoretic approaches.



\paragraph{Limitations}
The main limitation of our approach is the use of the approximation in \cref{eq:non_truncation_approximation}.
While this is necessary to keep all computations in the frequency domain, it leads to underestimation of the uncertainty of the non-causal part of the latent signal.
As such, our model ends up being misspecified even for the synthetic data we used in \cref{section:evaluation}.
In model comparison terms, we are operating in the ``\(\mathcal{M}\)-completed'' regime~\citep{bernardo_bayesian_2004}: the model is misspecified, but comparing model posterior probabilities is still useful.
Indeed, our approach showed good performance in \cref{section:evaluation}.
However, the effect of misspecification can be noticed in \cref{fig:detection_narrowband} in the high SNR regime.
Avoiding \cref{eq:non_truncation_approximation} while maintaining computational tractability would lead to a more powerful detection scheme in the high SNR-low snapshot regime.

\paragraph{Future Directions}
The strength of the Bayesian paradigm is that it is easy to encode prior knowledge and assumptions into the model.
Therefore, future directions would be to 
extend our model to incorporate colored noise~\citep{larocque_reversible_2002}, correlated sources, and non-uniform arrays.

{\appendices

\vspace{-2ex}
\section{Stripe-Decomposable Matrices}\label[appendix]{section:appendix_stripe}
We discuss the definition and properties of stripe-decomposable matrices.

\vspace{2ex}
\begin{definition}
We say a matrix \(\mA \in \mathbb{C}^{M L \times N L}\) is stripe decomposable with \(L \times L\) blocks if there exist a decomposition
\[
    \mA = \mPhi_M \, \mS \, \mPhi_{N}^{\herm} \; ,
\]
where, for any \(K \in \mathbb{N}_{>0}\), \(\mPhi_K \in \mathbb{C}^{K L \times K L}\) is some unitary matrix such that \(\mPhi_K \mPhi_K^{\herm} = \boldupright{I}_{KL}\), and the corresponding ``stripe matrix'' \(\mS \in \mathbb{C}^{M L \times N L}\) is structured as
\[
    \mS = \begin{bmatrix}
        \mS_{1,1} & \ldots & \mS_{1,N} \\
        \vdots & \ddots & \vdots  \\
        \mS_{M,1} & \ldots & \mS_{M,N} \\
    \end{bmatrix},
\]
where \(\mS_{i,j} \in \mathbb{C}^{L \times L}\) for \((i,j) \in [M] \times [N]\) is a diagonal matrix.
\end{definition}

\vspace{2ex}
\begin{proposition}\label{thm:eigendecomp_to_stripematrix}
    Let \(\mA \in \mathbb{C}^{M L \times N L}\) be a block-structured matrix such that
    \[
        \mA = \begin{bmatrix}
            \mA_{1,1} & \ldots & \mA_{1,L} \\
            \vdots & \ddots & \vdots \\
            \mA_{M,1} & \ldots & \mA_{M,L} \\
        \end{bmatrix},
    \]
    where each block \(\mA_{i,j} \in \mathbb{R}^{L \times L}\) admits a spectral decomposition \(\mA_{i,j} = \mP \mD_{i,j} \mP^{\herm}\) for \((i,j) \in [M] \times [N]\) with the same unitary basis matrix \(\mP\).
    Then \(\mA\) admits a stripe decomposition where the corresponding stripe matrix \(\mS\) structured such that \({[\mS]}_{i,j} = \mD_{i,j}\) and \(\mPhi_{M} = \mathrm{block\text{-}diagonal}_M\left( \mP, \ldots, \mP\right)\) and \(\mPhi_{N} = \mathrm{block\text{-}diagonal}_N\left( \mP, \ldots, \mP\right)\).
\end{proposition}
\begin{proof}
    It suffices to check that
    \begin{align}
        {\left[\mPhi_M \, \mS \, \mPhi_N^{\herm}\right]}_{i,j} 
        &= 
        \sum_{k=1}^M \sum_{l=1}^N {\left[\mPhi_M\right]}_{i,k} {\left[\mS\right]}_{k,l} {\left[\mPhi_N^{\herm}\right]}_{l,j}
        \nonumber
        \\
        &= 
        \sum_{k=1}^M \sum_{l=1}^N \mathds{1}_{i=k} \mathds{1}_{l=j} {\left[\mPhi_M\right]}_{i,k} {\left[\mS\right]}_{k,l} {\left[\mPhi_N^{\herm}\right]}_{l,j}
        \label{eq:spectraldecomp_stripe_proof_eq1}
        \\
        &= 
        {\left[\mPhi_M\right]}_{i,i} {\left[\mS\right]}_{i,j} {\left[\mPhi_N^{\herm}\right]}_{j,j}
        \nonumber
        \\
        &= 
        \mP \, \mD_{i,j} \, \mP^{\herm}
        = 
        \mA_{i,j} \; .
        \nonumber
    \end{align}
    \cref{eq:spectraldecomp_stripe_proof_eq1} is due to  \(\mPhi_{M}\) and \(\mPhi_N\) being block-diagonal.
\end{proof}

\begin{proposition}\label{thm:stripe_properties}
Let \(\mA = \mPhi_{M} \, \mS \, \mPhi_{N}^{\herm} \in \mathbb{C}^{M L \times N L}\), \(\mB = \mPhi_{M} \, \mT \, \mPhi_{N}^{\herm} \in \mathbb{C}^{M L \times N L} \), \(\mC = \mPhi_{N} \, \mU \, \mPhi_{K}^{\herm} \in \mathbb{C}^{N L \times K L} \), and \(\mD = \mPhi_{M} \mV \mPhi_M^{\herm} \in \mathbb{C}^{M L \times M L} \), be stripe-decomposable matrices with \(L \times L\) blocks, where \(\mD\) is further assumed to be invertible.
Then the following properties hold:
\begin{alignat*}{3}
  &\text{\(\bullet\) Hermitian Transpose:}\quad &&\mA^{\herm} &&= \,\mPhi_{N} \, \mS^{\herm} \, \mPhi_{M}^{\herm} \\
  &\text{\(\bullet\) Addition:}\quad &&\mA + \mB &&=  \,\mPhi_{M} \,\left(\mS + \mT \right) \,\mPhi_{N}^{\herm} \\
  &\text{\(\bullet\) Multiplication:}\quad  &&\mB \, \mC &&= \,\mPhi_{M} \, \mT \, \mU \, \mPhi_{K}^{\herm} \\
  &\text{\(\bullet\) Determinant:}\quad && \det\mD &&= \det \mV \\
  &\text{\(\bullet\) Inverse:}\quad &&\mD^{-1} &&= \,\mPhi_{M} \, \mV^{-1} \, \mPhi_{M}^{\herm}.
\end{alignat*}
\end{proposition}
\begin{proof}
    Transposition, addition, and multiplication are trivial.
    
    \paragraph{Proof of Determinant}
    From the properties of the determinant, we have \(\det\mD = \det \mPhi_M \det \mV \det \mPhi_M^{\herm} = \det \mV\), where the last equality follows from the fact that \(\mPhi_M^{-1} = \mPhi^{\herm}_M\).
    
    \paragraph{Proof of Inverse}
    Since \(\mD\) is invertible, the matrix \(\mD^{-1}\) such that \(\mD \mD^{-1} = \boldupright{I}_{NL}\) exists and is unique.
    Also, from the fact that \(\det \mD = \det \mV \neq 0\) it is evident that \(\mV\) is invertible such that \(\mV^{-1}\) uniquely exists.
    Now let \(\mE \triangleq \mPhi_{M} \mV^{-1} \mPhi_{M}^{\herm} \).
    It follows that 
    \begin{align*}
    \mD \mE
    &= 
    \left(\mPhi_{M} \mV \mPhi_{M}^{\herm} \right) \left(\mPhi_{M} \mV^{-1} \mPhi_{M}^{\herm} \right) 
    \\
    &= 
    \mPhi_{M} \mV \mV^{-1} \mPhi_{M}^{\herm}
    = 
    \mPhi_{M} \mPhi_{M}^{\herm}
    = 
    \boldupright{I}_{M L}.
    \end{align*}
    Therefore, it is clear that \(\mD^{-1} = \mE\).
    
\end{proof}

\begin{proposition}\label{thm:blockldl_is_ldl}
    Let \(\mS \in \mathbb{C}^{N L \times N L}\) be a positive definite stripe matrix with \(L \times L\) diagonal blocks.
    Applying \cref{alg:ldl} to \(\mS\) yields the block-LDL\(^{\top}\) decomposition 
    \[
         \mS = \mL \mD \mL^{\herm}.
    \]
    This is also the unique LDL\(^{\top}\) decomposition of \(\mS\).
\end{proposition}
\begin{proof}
    Since \(\mS\) is positive definite, its LDL\(^{\top}\) decomposition is unique.
    Now, since all the blocks of \(\mS\) are diagonal matrices and \(\mL\) has the same structure for the block-lower-triangular region, its upper-triangular, excluding the diagonal, is full of zeros.
    Therefore, \(\mL\) is already a lower-triangular matrix and similarly, \(\mD\) is already a diagonal matrix.
    Therefore, \(\mL \mD \mL^{\herm}\) is coincidentally the unique LDL\(^{\top}\) decomposition of \(\mS\).
\end{proof}

\vspace{-2ex}
\section{Derivation of \cref{eq:approximate_collapsed_likelihood}}\label[appendix]{section:approximate_collapsed_likelihood}

Recall that \cref{eq:collapsed_likelihood} is comprised as follows:
\begin{align*}
  {
    \underbrace{
    \mathrm{det}\left(
      \mP_{\bot}
    \right)
    }_{F_{\text{det}}}
  }^{-\nicefrac{1}{2}}
  {\big(
    \nicefrac{\alpha}{2}
    +
\underbrace{
    \vy^{\herm}
    \mP_{\bot}
    \vy
}_{F_{\text{quad}}}
  \big)}^{- M N / 2 + \beta}.
\end{align*}

\paragraph{Derivation of \(F_{\mathrm{quad}}\)}
From the definition of \(\mP_{\bot}\) in \cref{eq:pbot}, 
\begin{align}
    F_{\mathrm{quad}}
    =
    \vy^{\herm} \mP_{\bot} \vy
    =
    \vy^{\herm} \vy 
    +
    \vy^{\herm}
    \mA {\left(\mSigma^{-1} + \mA^{\herm} \mA \right)}^{-1} \mA^{\herm} \vy.
    \label{eq:quadratic_form}
\end{align}
From \cref{eq:system_stripe_identity}, we have
\begin{align}
    &{\left(\mSigma^{-1} + \mA^{\herm} \mA \right)}^{-1}
    \nonumber
    \\
    &\;=
    {\left({\left( \mPhi_k \, \mT \, \mPhi_k^{\herm} \right)}^{-1} + {\left(\mM_{M} \, \mPhi_M \, \mS \, \mPhi_k^{\herm}\right)}^{\herm} \mM_{M} \, \mPhi_M \, \mS \, \mPhi_k^{\herm} \right)}^{-1}
    \nonumber
    \\
    &\;=
    \mPhi_k
    {\left(  \mT^{-1}
    + 
    \mS^{\herm} \mPhi_M \mM_{M}^{\top} \mM_{M} \, \mPhi_M^{\herm} \, \mS  \right)}^{-1}
    \mPhi_k^{\herm} \; ,
    \nonumber
\shortintertext{where, applying the approximation in \cref{eq:non_truncation_approximation},} 
    &\;\approx
    \mPhi_k
    {\left(  \mT^{-1}
    + 
    \mS^{\herm} \mPhi_M \, \mPhi^{\herm}_M \, \mS  \right)}^{-1}
    \mPhi_k^{\herm}
    \nonumber
    \\
    &\;=
    \mPhi_k
    {\left( \mT^{-1} +  \mS^{\herm} \, \mS  \right)}^{-1}
    \mPhi_k^{\herm} \; .
    \nonumber
\end{align}
Therefore,
\begin{align}
    &F_{\mathrm{quad}}
    \;\approx
    \vy^{\herm} \mA \,
    \mPhi_k
    {\left( \mT^{-1} +  \mS^{\herm} \, \mS  \right)}^{-1}
    \mPhi_k^{\herm} \,
    \mA^{\herm}
    \vy 
    \nonumber
    \\
    &\;=
    \vy^{\herm} \, \left( \mM_{M} \, \mPhi_M \, \mS \, \mPhi_k^{\herm} \right) \, 
    \mPhi_k
    {\left( \mT^{-1} +  \mS^{\herm} \, \mS  \right)}^{-1}
    \nonumber
    \\
    &\qquad\qquad
    \times \mPhi_k^{\herm} \,
    {\left( \mM_{M} \, \mPhi_M \, \mS \, \mPhi_k^{\herm} \right)}^{\herm}
    \vy 
    \nonumber
    \\
    &\;=
    \vy^{\herm} \,  
    \mM_{M} \, \mPhi_M \, \mS \, 
    \left( \mPhi_k^{\herm}  \mPhi_k \right)
    {\left( \mT^{-1} +  \mS^{\herm} \, \mS  \right)}^{-1}
    \nonumber
    \\
    &\qquad\qquad
    \times
    \left( \mPhi_k^{\herm} \, \mPhi_k \right)
    \mS^{\herm} \mPhi_M^{\herm} \mM_{M}^{\top}
    \vy 
    \nonumber
    \\
    &\;=
    \vy^{\herm} \, \mM_{M} \, \mPhi_M \, \mS \, 
    {\left( \mT^{-1} +  \mS^{\herm} \, \mS  \right)}^{-1}
    \mS^{\herm} \, \mPhi^{\herm}_M \mM_{M}^{\top} \vy 
    \nonumber
    \\
    &\;=
    \vy_{\mathcal{F}}^{\herm} \mS \, 
    {\left( \mT^{-1} +  \mS^{\herm} \, \mS  \right)}^{-1}
    \mS^{\herm} \, \vy_{\mathcal{F}} \; .
    \nonumber
\end{align}

\paragraph{Derivation of \(F_{\mathrm{det}}\)}
From the matrix determinant lemma and the definition of \(\mP_{\bot}\) in in \cref{eq:pbot}, 
\begin{align}
  F_{\mathrm{det}}
  &=
  \mathrm{det}\left( \mA \mSigma \mA^{\herm}  + \mathbf{I} \right)
    \nonumber
  \\
  &=
  \mathrm{det}\left( \mSigma^{-1} \right) / \mathrm{det}\left( \mSigma^{-1} + \mA^{\herm} \mA \right)
    \nonumber
  \\
  &\;=
  \mathrm{det}\left( \mSigma^{-1} \right) /
  {\mathrm{det}\left( 
    \mPhi_k \left(
    \mT^{-1}  + \, \mS^{\herm} \mM_{M}^{\top} \mM_{M} \, \mS 
    \right)
    \mPhi_k^{\herm}
  \right)} \; ,
    \nonumber
\shortintertext{and applying the approximation in \cref{eq:non_truncation_approximation},} 
  &\;\approx
  \mathrm{det}\left( \mSigma^{-1} \right) /
  {\mathrm{det}\left( 
    \mPhi_k \left(
    \mT^{-1}  + \, \mS^{\herm} \mS 
    \right)
    \mPhi_k^{\herm}
  \right)}
    \nonumber
  \\
  &\;=
  \mathrm{det}\left( \mT^{-1} \right) /
  {\mathrm{det}\left( 
    \mT^{-1}  + \, \mS^{\herm} \mS 
  \right)} \; .
    \nonumber
\end{align}

}

\section*{Acknowledgments}
The authors sincerely thank Philippe Gagnon for comments on advanced RJMCMC proposals; Samuel Power and Trevor Campbell for discussions on slice samplers;  Miguel Biron-Lattes and Dootika Vats for advice on estimating the autocorrelation of non-reversible Markov chains; Ian Waudby-Smith for discussions on hypothesis testing.

This work was supported by the UK Defence Science and Technology Laboratories (Dstl, Grant no. 1000143726) as part of Project BLUE under the UK MoD University Defence Research Collaboration (UDRC) in Signal Processing.
K. Kim was also partly supported by a gift from AWS AI to Penn Engineering's ASSET Center for Trustworthy AI.

\bibliographystyle{IEEEtranN}
\bibliography{bstcontrol,IEEEabrv,references}



\begin{IEEEbiography}[{\includegraphics[width=1in,height=1.25in,clip,keepaspectratio]{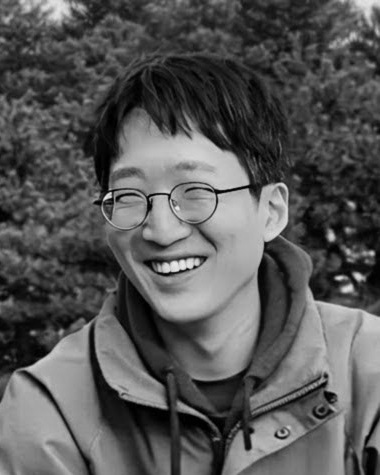}}]{Kyurae~Kim}
(Member, IEEE) received the B.S. degree from the Dept. of Electronics Engineering, Sogang University, Seoul, South Korea, in 2021.
He is working towards his Ph.D. with the Dept. of Computer and Information Sciences, University of Pennsylvania, Philadelphia, U.S.

From 2021 to 2022, he was a Research Associate with the Department of Electrical Engineering and Electronics, University of Liverpool, Liverpool, U.K. 
His research interests include Bayesian inference methods, stochastic optimization, and their signal-processing applications. 

Kim is a member of the Association for Computing Machinery (ACM) and the International Society for Bayesian Analysis (ISBA).
\end{IEEEbiography}

\vspace{-5ex}
\begin{IEEEbiography}[{\includegraphics[width=1in,height=1.25in,clip,keepaspectratio]{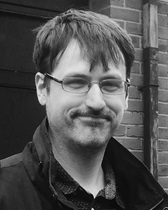}}]{Philip~T.~Clemson}
received the M.Phys. and Ph.D. degrees in physics from Lancaster University, Lancaster, U.K., in 2009 and 2013 respectively.

From 2019 to 2023, he was a research associate with the Dept. of Electrical Engineering and Electronics, University of Liverpool, Liverpool.
He is now a research associate at the Physics Department, Lancaster University, Lancaster, U.K. His research interests include time series analysis of non-autonomous dynamical systems and their application to biomedical data, as well as Bayesian approaches applied to signal processing.
\end{IEEEbiography}

\vspace{-5ex}
\begin{IEEEbiography}[{\includegraphics[width=1in,height=1.25in,clip,keepaspectratio]{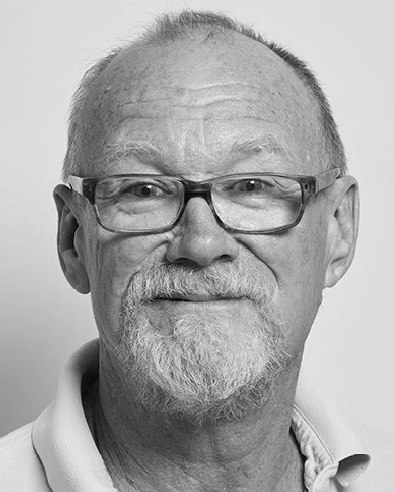}}]{James P. Reilly}
(Life Member, IEEE) 
received the B.A.Sc. degree from the University of Waterloo,
Waterloo, ON, Canada, in 1973, and the M.Eng. and Ph.D. degrees from McMaster University, Hamilton, ON, Canada, in 1977 and 1980, respectively, all in electrical engineering. 

He was employed in the telecommunications industry for a total of 7 years and was then appointed to the Dept. of Elec. \& Comp. Eng. at McMaster University in 1985. 
He has been a visiting academic at the University of Canterbury, New Zealand, and the University of Melbourne,
Australia. 
His research interests include several aspects of signal processing, specifically machine learning, EEG signal analysis, Bayesian methods, blind signal separation, blind identification, and array signal processing. He has contributed pioneering work in the use of machine learning and signal processing to the characterization of various forms of brain disorders.
\end{IEEEbiography}

\begin{IEEEbiography}[{\includegraphics[width=1in,height=1.25in,clip,keepaspectratio]{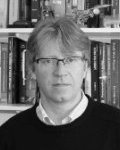}}]{Jason~F.~Ralph}
received the B.Sc. degree in physics with mathematics from the University of Southampton, Southampton, U.K., in 1989, and the D.Phil. degree from the University of Sussex, Brighton, U.K., in 1993. 

He is currently a Professor with the Dept. of Electrical Engineering and Electronics, University of Liverpool, Liverpool, U.K., and served as the Head of the Department from 2012 to 2015.
His research interests include quantum technologies, guidance and navigation, and target-tracking algorithms.
\end{IEEEbiography}

\begin{IEEEbiography}[{\includegraphics[width=1in,height=1.25in,clip,keepaspectratio]{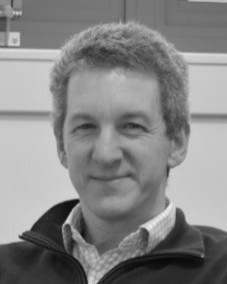}}]{Simon~Maskell}
(Member, IEEE) received the M.A., M.Eng., and Ph.D. degrees in engineering from the University of Cambridge, Cambridge, U.K., in 1998, 1999, and 2003, respectively.

Before 2013, he was a Technical Manager of command, control and information systems with QinetiQ, U.K. Since 2013, he has been a Professor of Autonomous Systems with the University of Liverpool, Liverpool, U.K. 
His research interests include Bayesian inference applied to signal processing, multitarget tracking, data fusion, and decision support with particular emphasis on the application of sequential Monte Carlo methods in challenging data science contexts. 
He was an Associate Editor for \textsc{IEEE Transactions of Aerospace and Electronic Systems} and \textsc{IEEE Signal Processing Letters} and is now a Dstl/Royal Academy of Engineering Chair in Information Fusion.

Prof. Maskell also served as President of the International Society of Information Fusion (ISIF) and is now the Secretary.
\end{IEEEbiography}

\vfill

\end{document}